\documentclass[journal=jctcce,manuscript=article,layout=traditional]{achemso}

\usepackage{graphicx}
\usepackage{dcolumn}%
\usepackage{bm}%
\usepackage{latexsym}
\usepackage{amsmath}
\usepackage{amssymb}
\usepackage{amsfonts}
\usepackage{xcolor}
\usepackage{braket}
\usepackage{comment}
\usepackage{booktabs}
\usepackage{xspace}
\usepackage{multirow}
\usepackage[colorlinks,linkcolor=blue,citecolor=blue,urlcolor=blue]{hyperref}
\def\vsqrt{v^{1/2}}

\def\br{\mathbf{r}}

\def\h2o{\mathrm{H}_2\mathrm{O}}

\def\icomp{\mathrm{i}}
\def\e{\epsilon}
\def\w{\omega}
\def\W{\Omega}

\DeclareMathOperator{\im}{Im}

\newcommand{\pph}{p-(p-h)\xspace}
\newcommand{\hhp}{h-(h-p)\xspace}
\newcommand{\aug}{$GW$+2SOSEX-aug\xspace}
\newcommand{\psd}{$GW$+2SOSEX-psd\xspace}

\newcommand{\subfigimg}[3][,]{%
  \setbox1=\hbox{\includegraphics[#1]{#3}}
  \leavevmode\rlap{\usebox1}
  \rlap{\hspace*{0pt}\raisebox{\dimexpr\ht1-0\baselineskip}{#2}}
  \phantom{\usebox1}
}

\DeclareUnicodeCharacter{2212}{-}
\DeclareUnicodeCharacter{2061}{}

\title{$GW$+2SOSEX self-energy made positive semi-definite}

\author{Fabien Bruneval}
\email{fabien.bruneval@cea.fr}
\affiliation{Université Paris-Saclay, CEA, Service de recherche en Corrosion et Comportement des Matériaux, SRMP, 91191 Gif-sur-Yvette, France}

\author{Arno Förster}
\email{a.t.l.foerster@vu.nl}
\affiliation{Theoretical Chemistry, Vrije Universiteit, De Boelelaan 1108, 1081 HZ Amsterdam, The Netherlands}

\author{Yaroslav Pavlyukh}
\email{yaroslav.pavlyukh@pwr.edu.pl}
\affiliation{Institute of Theoretical Physics, Faculty of Fundamental Problems of Technology, Wroclaw University of Science and Technology, 50370 Wroclaw, Poland}

\date{\today}

\begin{document}

\begin{abstract}
The formulation of vertex corrections beyond the $GW$ approximation within the framework of perturbation theory is a subtle and challenging task, which accounts for the wide variety of schemes proposed over the years. Exact self-energies are required to satisfy the mathematical condition of positive semi-definiteness. The $GW$ self-energy fulfills this property, but the vast majority of the vertex-corrected self-energy approximations do not.
In this study, we devise a positive semi-definite extension to the $GW$+2SOSEX self-energy that we name $GW$+2SOSEX-psd.
To reach this goal, we demonstrate the cancellation of the bare energy poles that are contained in the fully dynamic second-order in $W$ self-energy ($G3W2$).
We then demonstrate on molecular examples the correct positive semi-definiteness of the proposed self-energy approximation and its good accuracy in predicting accurate quasiparticle energies for valence and core states.
\end{abstract}

\begin{tocentry}
\includegraphics[scale=0.39]{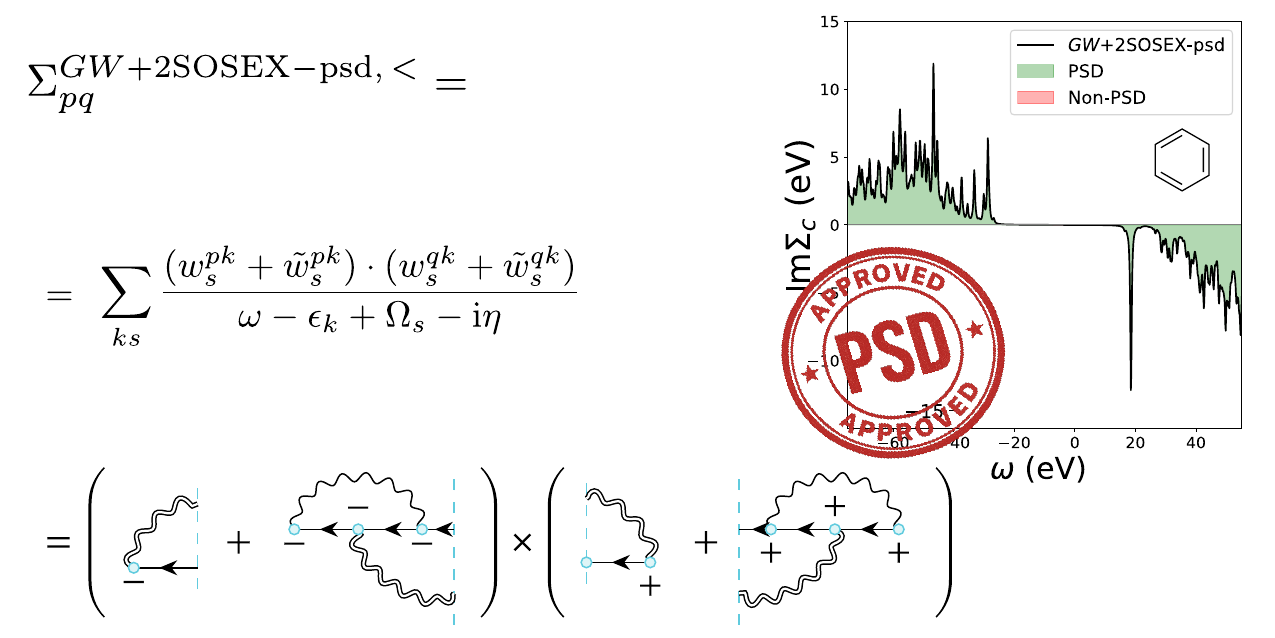}
\end{tocentry}

\maketitle

\section{Introduction}
\label{sec:intro}

Over the years, Green's functions have proven to be insightful and efficient tools for addressing the many-electron problem in both extended and finite systems. \cite{cederbaum_book1977,strinati_rnc1988,mahan_book,stefanucci_nonequilibrium_2013}.
In this approach, the central unknown object is the self-energy operator $\Sigma$
that governs
the equation-of-motion of the Green's function (GF) $G$.
On this road, Hedin's $GW$ approximation to the self-energy \cite{hedin_pr1965}
represents a unique milestone:
it is an appealing combination of simplicity and accuracy
\cite{reining_wires2018,golze_fchem2019, Marie2024b}.
In contrast with regular perturbation theory based on the bare Coulomb interaction $v$
\cite{moller_pr1934, cederbaum_tca1973},
the $GW$ approximation is obtained as the first-order expansion in terms of the 
screened Coulomb interaction $W$.

Producing approximations beyond the $GW$ term, often referred to as
``vertex corrections'', is not a straightforward task.
An order-by-order expansion in terms of $v$ is unique and well-defined,
as illustrated by the early results by Cederbaum
\cite{cederbaum_tca1973,cederbaum_jpb1975,cederbaum_book1977} with
perturbation theory to second order (PT2), third order (PT3) etc. Electron screening effects represent a problem on their own.
Using the dynamically screened interaction $W(\omega)$ systematically, as a building block of perturbative expansions, is much more difficult, both conceptually\cite{ holm_prb1998, stefanucci_prb2014} and numerically.
For this reason, following some early attempts in the 90s\cite{Mahan1989, delsole_prb1994, Shirley1996, DeGroot1996, Schindlmayr1998} we have witnessed a blooming of vertex correction proposals for extended\cite{Shishkin2007, gruneis_prl2014, Kutepov2016, Schmidt2017,Ma2019a, Pavlyukh2020, Tal2021, Cunningham2023, Rohlfing2023} and finite\cite{romaniello_jchemphys2009, ren_prb2015, Kuwahara2016, Vlcek2019, lewis_jctc2019,Ma2019a,bruneval_fchem2021, Mejuto-Zaera2022a, Loos2024, bruneval_jctc2024, Vacondio2024, foerster_jpcl2024,patterson_jctc2024, Patterson2024a, Forster2025} systems. 
These many propositions come from the difficult task of truncating the 3-point vertex equation, and many different choices can be made.\cite{strinati_rnc1988} 
The vertex can be included in $W$ only,\cite{Shishkin2007, Schmidt2017, Cunningham2023, lewis_jctc2019, bruneval_fchem2021} in the self-energy only,\cite{Romaniello2012, ren_prb2015, Pavlyukh2020, bruneval_jctc2024} or in both.\cite{romaniello_jchemphys2009, gruneis_prl2014, Kutepov2016, Kuwahara2016, Vlcek2019, Ma2019a, Tal2021, Mejuto-Zaera2022a, Rohlfing2023, Vacondio2024, foerster_jpcl2024,patterson_jctc2024, Patterson2024a, Forster2025} At least for molecules, many works have shown the latter strategy to be rather successful.\cite{Vlcek2019,Vacondio2024, foerster_jpcl2024, patterson_jctc2024, Patterson2024a, Forster2025} These advanced strategies are beyond the scope of this work, and we only consider a vertex correction in the self-energy, keeping $W$ at the level of random phase approximation (RPA).

Very often, the focus is placed on the main peaks of the imaginary part of the Green's function $G(\w)$, 
the so-called spectral function $A(\w)$:
\begin{align}
    G(\w)=\int_{\mathcal{C}} \frac{d\w'}{2\pi}\frac{A(\w')}{\w-\w'},
\end{align}
where the integration contour $\mathcal{C}$ goes infinitesimally above and below the real axis for $\w'\le\mu$, and $\w'>\mu$, respectively, with $\mu$ being the chemical potential of the system. These peaks mark the ionization or affinity energies of electrons in an interacting electron system and can be compared to results of actual photoemission and inverse-photoemission experiments. When the peaks are well defined, they are named quasiparticle ($qp$) peaks. However, the spectral function often exhibits additional complexity beyond the main peaks, in the form of the so-called satellite structures. They are well-known spectral features in the homogeneous electron gas~\cite{holm_self-consistent_1997, takada_low-energy_2024}, bulk materials~\cite{caruso_band_2015, zhou_cumulant_2018, riley_crossover_2018}, and molecules~\cite{Mejuto-Zaera2021, Loos2024, kocklauner_gw_2025}. 

Both effects, the renormalization of the $qp$ peaks and the emergence of satellites in the spectral function, is encoded in the electron self-energy (SE) $\Sigma(\w)$, which can, in turn, be recast in terms of its spectral function $\Gamma(\omega)$:
\begin{align}
\Sigma(\w)&=\Sigma^x+\Sigma^c(\w)=\Sigma^x+\int_{\mathcal{C}} \frac{d\w'}{2\pi}\frac{\Gamma(\w')}{\w-\w'},
\label{eq:lehmann}
\end{align}
where $\Sigma^x$ is the frequency-independent exchange part of the self-energy, 
and $\Sigma^c(\omega)$ is the correlation part. The function $\Gamma(\omega)$ is sometimes referred to as a rate function due to its physical interpretation—it describes the rate at which various scattering processes influence the many-body state of the system following the removal or addition of a particle (electron). A fundamental requirement is that this rate must be non-negative for all energy transfers and for all combinations of initial and final states. This leads to the mathematical requirement that $\Gamma(\omega)$ must be a positive semi-definite (PSD) matrix ($\Gamma(\omega) \succeq 0$) admitting a Lehmann (i.e., sum over states) representation~\cite{stefanucci_prb2014,uimonen_prb2015,pavlyukh_prl2016}. For the time-ordered self-energy, the PSD condition translates to the constraints:
\begin{equation}
\label{eq:psd}
\left\{\begin{aligned}
  \im{\langle x| \Sigma^c(\omega)}| x\rangle \ge 0 &\text{ when } \omega \le \mu, \\
  \im{\langle x|\Sigma^c(\omega)}| x\rangle \le 0 & \text{ when } \omega > \mu \\
 \end{aligned}\right.
\end{equation}
for any state $x$ onto which we project. A PSD self-energy implies two key consequences: (i) it gives rise to a PSD spectral function $A(\w)\succeq0$---ensuring a positive electron density as expected, and (ii) it preserves the correct causal structure: the self-energy is derived from $\Gamma(\w)$ via the Kramers-Kr\"{o}nig relation (Eq.~\ref{eq:lehmann}), and $\Gamma(\w)$ is determined as the imaginary part of $\Sigma(\w)$. The fact that the exact self-energy admits a Lehmann representation has been known since the early work of Winter~\cite{winter_study_1972}. However, it was largely overlooked that, starting from third order in the bare interaction, some self-energy diagrams violate this property~\cite{cederbaum_jchemphys1975}, a point that only gained attention through early calculations for the homogeneous electron gas by Minnhagen~\cite{minnhagen_jpc1974} and through Almbladh’s analysis of photoemission processes~\cite{almbladh_theory_1985}.

\begin{figure*}[hbt!]
 \centering
  \begin{tabular}{@{}p{0.45\linewidth}@{\qquad\qquad}p{0.45\linewidth}@{}}
    \subfigimg[width=\linewidth]{a)}{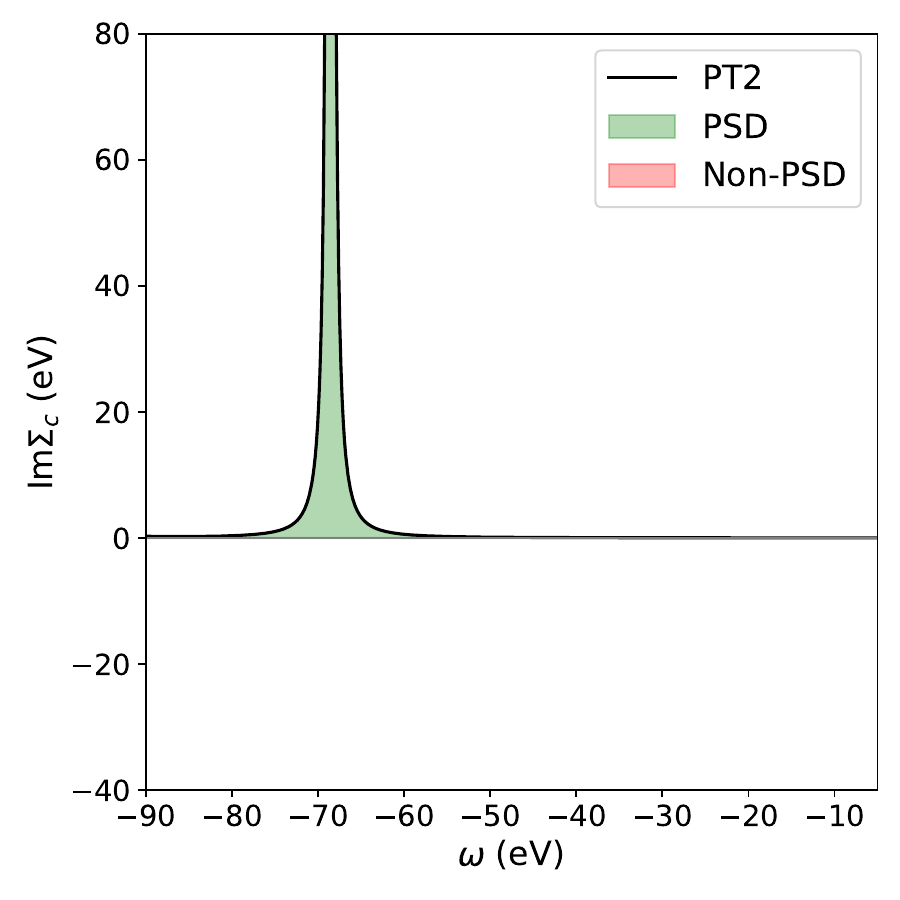} &
    \subfigimg[width=\linewidth]{b)}{figs/ne_GW} \\
    \subfigimg[width=\linewidth]{c)}{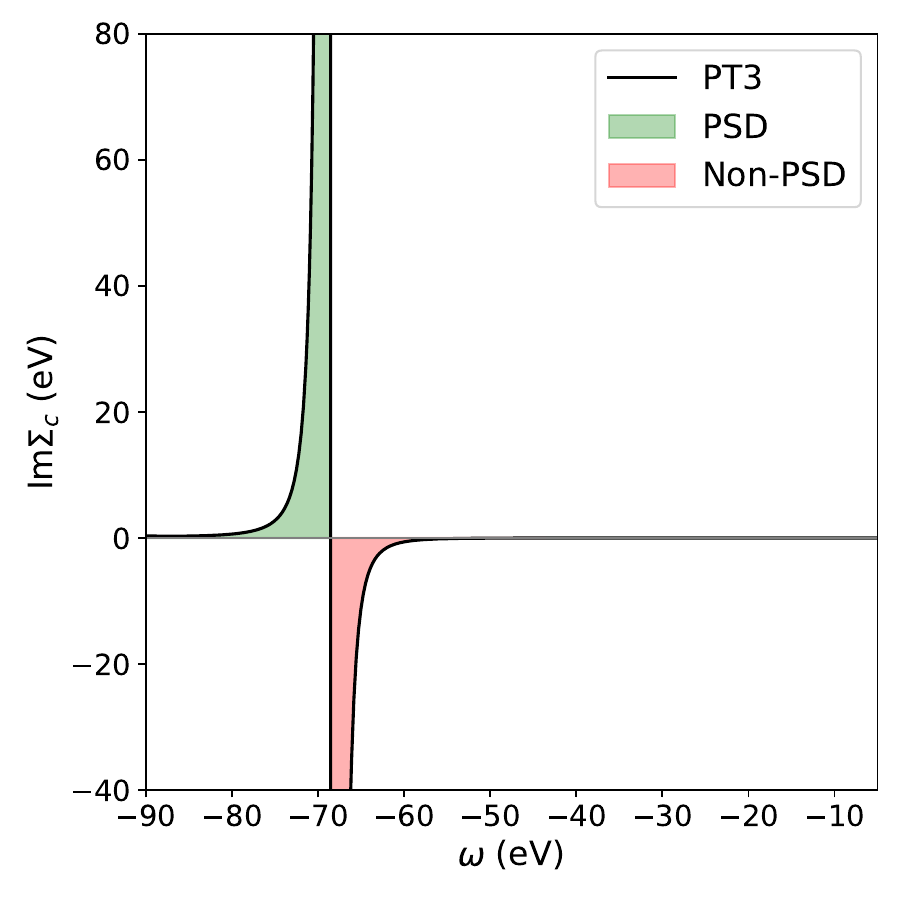} &
    \subfigimg[width=\linewidth]{d)}{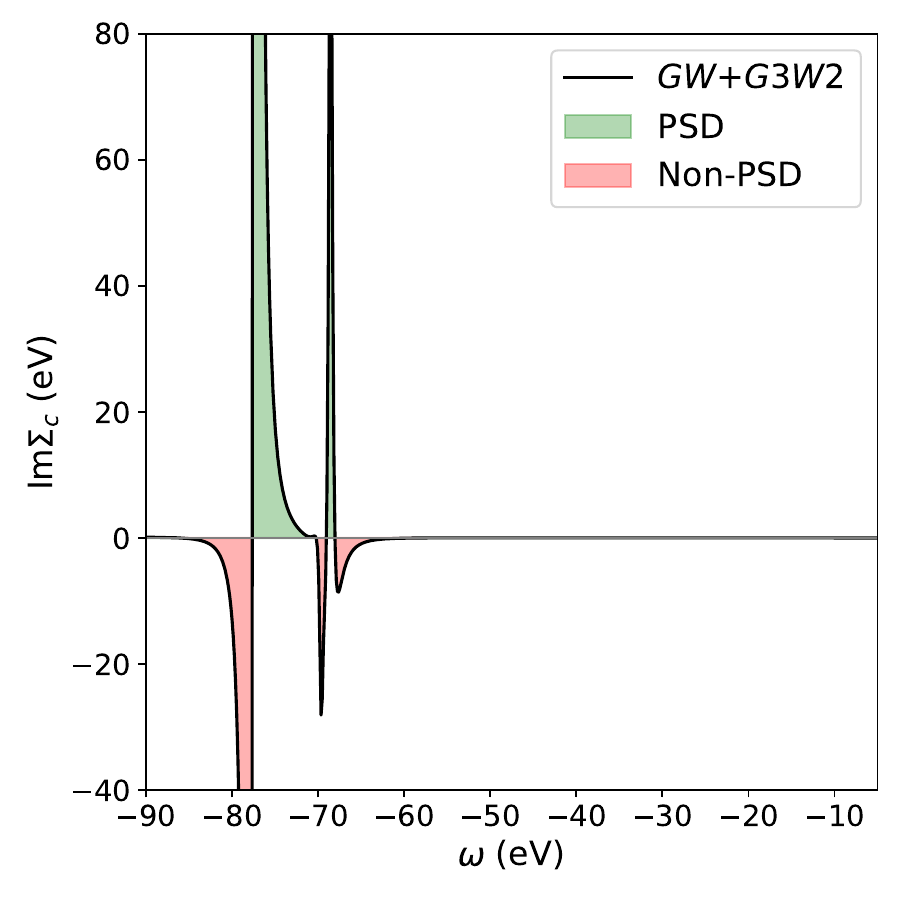} 
  \end{tabular}
\caption{
Imaginary part of selected time-ordered correlation self-energies for the Ne atom.
We plot the HOMO diagonal expectation value $\mathrm{Im} \langle \mathrm{HOMO} | \Sigma^c(\omega) | \mathrm{HOMO} \rangle$ for PT2, $GW$ in the first row, and PT3, $GW+G3W2$ in the second row. 
}
\label{fig:neon1}
\end{figure*}

Let us check out whether common approximations fulfill the PSD property
with the example of a single neon atom in Fig.~\ref{fig:neon1}.
The two simplest approximations are presented in the upper row:  a $v$-based expansion (PT2) in panel (a)  and a $W$-based expansion ($GW$) in panel (b). Both approximations appear as PSD, and we will recall in this article that it is a general mathematical result.
However, when trying to include higher-order terms in panel (c) with PT3 (Ref.~\citenum{cederbaum_tca1973}) or in panel (d) with the fully dynamic $G3W2$ term (Refs.~\citenum{minnhagen_jpc1974,bruneval_jctc2024}), large negative parts show up.

\begin{figure*}[t]
 \centering
  \begin{tabular}{@{}p{0.5\linewidth}@{\qquad}p{0.35\linewidth}@{}}
    \subfigimg[scale = 1.2]{a)}{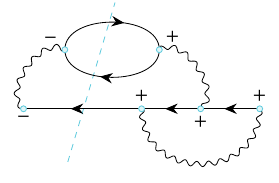} &
    \raisebox{1.1cm}{\subfigimg[scale = 1.2]{b)}{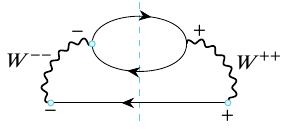}}\\
    \raisebox{0.7cm}{\subfigimg[scale = 1.2]{c)}{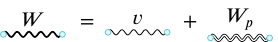}}& 
    \subfigimg[scale = 1.2]{d)}{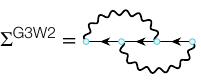}\\[0.5cm]
    \multicolumn{2}{c}{\subfigimg[scale = 1.2]{e)}{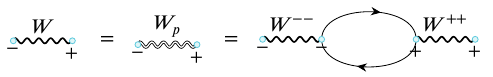}\hfill}
  \end{tabular}
\caption{
Diagrammatic proof why some diagrams of the PT3 approximation are not PSD (a), whereas the $GW$ approximation is PSD (b). Decomposition of the screened interaction $W$ into the bare Coulomb $v$ and a polarizable part $W_p$, viz. Eq.~\eqref{eq:wp} (c). The full second-order in $W$ SE diagram (d). Representation of the lesser RPA $W$ component in terms of the time-ordered ($W^{--}$), electron-hole bubble, and anti-time-ordered $W^{++}$ components (e). 
}
\label{diags:SE}
\end{figure*}

These observations can be readily rationalized using the method introduced in Ref.\citenum{stefanucci_prb2014}, based on the non-equilibrium Green's function (NEGF) formalism, as illustrated in Fig.~\ref{diags:SE}, panels (a) and (b). Below, we sketch the idea, and we will provide more detailed arguments in the subsequent sections. To this end, consider the partitions (diagrams with pluses and minuses assigned to vertices indicating the branches of the Keldysh contour on which the corresponding time arguments lie) of the lesser self-energy (SE). Since there can be no isolated islands of pluses or minuses, the diagrams naturally split into two half-diagrams (as indicated by dashed lines), with time arguments residing exclusively on either the forward (minus) or backward (plus) branches. It is important to note then, that one half-diagram in panel (a) contains one interacting line, whereas the other contains two interacting lines. Conversely, SE diagrams are obtained by gluing half-diagrams (possibly with permutation of the dangling lines). Only when the plus and minus half-diagrams are equivalent can the glued expression be recast as a sum of squares (in the form of the Fermi golden rule), which is then manifestly positive definite. For the diagram in panel (a), this condition is not fulfilled due to the differing number of interaction lines, and hence the diagram is not PSD. Remedying this would require the addition of two more diagrams of second and fourth order in the bare interaction.

In contrast, the $GW$ diagram in panel (b) is PSD because its two constituent half-diagrams are topologically equivalent. A subtle point here is the representation of its constituent lesser screened interaction in terms of the time-ordered ($W^{--}$) and anti-time-ordered ($W^{++}$) components, as depicted in panel (e). We will present a more explicit positivity proof in Sec.~\ref{sec:gw}.

We also note that the full $G3W2$ diagram shown in Fig.~\ref{diags:SE}, panel (d), contains the non-PSD diagram from panel (a) as a subset. This suggests (we are cautious here because one non-PSD part can, in principle, be compensated by another, more positive part) that the full diagram may also fail to be PSD, a conclusion supported by the numerical results shown in Fig.~\ref{fig:neon1}, panel (d). We will return to this example in Sec.~\ref{sec:G3W2} explicitly demonstrating the non-positivity of its specific term. This is a serious shortcoming for the $G3W2$ approximation that calls for improvements.

In this work, we analyze the PSD behavior of approximations in the $G3W2$ family,
namely second-order exchange (SOX), second-order screened exchange (SOSEX) \cite{Romaniello2012, ren_prb2015},
complete second-order screened exchange (2SOSEX) \cite{bruneval_jctc2024}, and full $G3W2$.
These approximations are obtained by replacing one or two $W$ in $G3W2$ by the static $v$.
This will allow us to propose a PSD-compliant self-energy approximation
that comprises the 2SOSEX terms.
In doing so, we identify the role of poles in the self-energy
that combine 3 bare $qp$ energies and show how these contributions
can be rigorously discarded. The proposed self-energy approximations will be tested on finite systems for valence and core excitations using the established benchmarks
GW100 \cite{vansetten_jctc2015} and CORE65 \cite{golze_jpcl2020}.

\section{Diagnosing PT2, $GW$, $G3W2$ self-energies}
\label{sec:diagnose}

Since the article focuses on finite systems, we will employ notations commonly used in quantum chemistry.
Let us state that we will limit our discussion to a non-self-consistent Green's function
using a so-called one-shot approach and 
to the RPA to $W$.
Hartree atomic units are used everywhere in the text.


\subsection{Notations}

We use the usual convention where the indices $a, b, c$ run over virtual molecular orbitals (MO), $i, j, k$ over occupied MO, and $m, n, p, q$ over all the MO.
$\eta$ stands for a small positive real number that enforces the proper location of the poles in the complex plane: above the real axis for occupied states ($\epsilon_i < \mu$) and below the real axis for virtual states ($\epsilon_a > \mu$).
Note that we assume real wavefunctions without loss of generality for finite systems.

With this, the non-interacting time-ordered Green's function $G$ that comes from a previous generalized Kohn-Sham calculation is spin-diagonal and reads
\begin{equation}
  \label{eq:g0}
  G(\br,\br',\omega) =
    \sum_i^\mathrm{occ}
     \frac{\varphi_i(\br)   \varphi_i(\br^\prime)}
     {\omega - \epsilon_i -\icomp \eta }
+ \sum_a^\mathrm{virt}
     \frac{\varphi_a(\br)   \varphi_a(\br^\prime)}
     {\omega - \epsilon_a +\icomp \eta } .
\end{equation}
Along with the time-ordered  $G\equiv G^{--}$ and the anti-time-ordered $G^{++}(\w)=[G^{--}(\w)]^*$ components, the lesser $ G^<\equiv G^{-+}$ and greater  $G^>\equiv G^{+-}$ components of GF will be useful:
\begin{subequations}
\label{def:G:lessgtr}
\begin{align}
    G^<(\br,\br',\omega)&
    =\sum_{i} \varphi_i(\br) G^<_{i}(\omega)\varphi_{i}(\br^\prime)
    =2\pi \icomp \sum_i \varphi_i(\br)   \varphi_i(\br^\prime)\delta(\w-\epsilon_i),\\
    G^>(\br,\br',\omega)&
    =\sum_{a} \varphi_a(\br) G^<_{a}(\omega)\varphi_{a}(\br^\prime)
    =-2\pi \icomp \sum_a \varphi_a(\br) \varphi_a(\br^\prime)\delta(\w-\epsilon_a).
\end{align}
\end{subequations}
It is common practice to split the screened Coulomb interaction into a static part $v$
and a polarizable part $W_p(\omega)$ (see Fig.~\ref{diags:SE} panel c for diagrammatic representation):
\begin{equation}
\label{eq:wp}
 W(\br, \br^\prime, \omega) =  v(\br, \br^\prime) + W_p(\br, \br^\prime, \omega) .
\end{equation}
The polarizable part has a Lehmann representation:
\begin{equation}
\label{eq:wspec}
  W_p(\br,\br^\prime,\omega) =
    \sum_s w_s(\br) w_s(\br^\prime) 
       \left[ \frac{1}{\omega - \Omega_s + \icomp \eta}
             -\frac{1}{\omega + \Omega_s - \icomp \eta}
       \right]  \;,
\end{equation}
where the Lehmann amplitudes $w_s(\br)$ and the system's neutral excitation energies $\Omega_s$ can be obtained from the solution of Casida's equation
(or Bethe--Salpeter equation) within the RPA\cite{casida_book1995, casida_time-dependent_2009}. Note that $w_s$ incorporates the spin degeneracy factor. Here and in the following the index $s$ (and later also $t$) runs over the pairs of states
(occupied $\times$ virtual). Along with the time-ordered component of $W\equiv W^{--}$, the lesser and greater components of the screened interaction will be useful:
\begin{subequations}
\label{def:W:lessgtr}
\begin{align}
    W^<(\br,\br^\prime,\omega)&=\sum_{s} w_s(\br) W_{s}^<(\w) w_{s}(\br^\prime) =-2\pi\icomp \sum_s w_s(\br) w_s(\br^\prime)\delta(\w+\Omega_s),\\
    W^>(\br,\br^\prime,\omega)&=\sum_{s} w_s(\br) W_{s}^>(\w) w_{s}(\br^\prime) =-2\pi\icomp\sum_s w_s(\br) w_s(\br^\prime)\delta(\w-\Omega_s).
\end{align}
\end{subequations}
From the orbital/neutral excitation basis representation of $G$ (Eq.~\ref{def:G:lessgtr}) and $W$ (Eq.~\ref{def:W:lessgtr}), it is evident that the corresponding matrices are diagonal. 
This is advantageous, as it renders the diagrammatic technique akin to that used for periodic systems, where each propagator carries a momentum. However, in contrast to systems such as the electron gas, where performing the spatial integration over each vertex enforces momentum conservation, the integration here yields a nontrivial matrix element, given by the following integral:
\begin{equation}
 w_s^{pq} = \int d\br \varphi_p(\br)\varphi_q(\br) w_s(\br)  .
\end{equation}

In the following, we will work on 
the correlation part of self-energy that is dynamic $\Sigma^c(\omega)$.
The static part accounting for the exchange 
(and possibly for the corrections to the one-body reduced density-matrix
\cite{cederbaum_book1977, bruneval_prb2019})
will be added.

\subsection{PT2 is PSD\label{sec:pt2}}
\begin{figure*}
\includegraphics[scale=1.2]{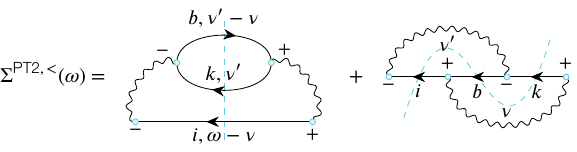} 
\caption{Diagrammatic representation of the two terms contributing to the PT2 approximation. Vertices are labeled by $-$/$+$ depending on the branch of Keldysh contour they reside on. Since GFs are diagonal in orbital basis, each fermionic line is labeled by the orbital index. Frequencies (e.g. $\nu$, $\nu'$) associated with some propagators are also indicated; the others follow from the energy conservation at each vertex.
}
\label{diag:pt2}
\end{figure*}
The second-order approximation to the self-energy, also named GF2 or the second Born approximation,
is the sum of two terms (Fig.~\ref{diag:pt2}): the 1-ring term and the SOX term \cite{szabo_book}.

For convenience, we split the terms according to the location of the pole with respect to the real axis of the complex plane.
Notation $vov$ stands for virtual-occupied-virtual, which will result in a pole in the fourth quadrant, whereas $ovo$ for occupied-virtual-occupied generates a pole in the second quadrant.

These terms read 
\begin{subequations}
\label{eq:1-ring}
\begin{equation}
\label{eq:1-ring:o}
\Sigma_{pq}^{\mathrm{1-ring},\, ovo}(\omega)
  = \sum_{ibk} 
   \frac{ 2 (pi|bk) \cdot  (qi|bk) }
        {\omega -\epsilon_i  + \epsilon_b -\epsilon_k   - \icomp \eta }
\end{equation}
\begin{equation}
\Sigma_{pq}^{\mathrm{1-ring},\, vov}(\omega) 
 = \sum_{a j c} 
     \frac{ 2 (pa|jc) \cdot (qa|jc) } 
          {\omega -\epsilon_a +\epsilon_j -\epsilon_c + \icomp \eta }
\end{equation}
\end{subequations}
and 
\begin{subequations}
\label{eq:SOX}
\begin{equation}
\label{eq:soxovo}
\Sigma_{pq}^{\mathrm{SOX},\, ovo}(\omega)
  = -\sum_{ibk} 
     \frac{ (pi|bk) \cdot (qk|bi) }
        {\omega -\epsilon_i  + \epsilon_b -\epsilon_k   - \icomp \eta }
\end{equation}

\begin{equation}
\label{eq:soxvov}
\Sigma_{pq}^{\mathrm{SOX},\, vov}(\omega) 
 = -\sum_{a j c} 
     \frac{ (pa|jc) \cdot (qc|ja) } 
          {\omega -\epsilon_a +\epsilon_j -\epsilon_c + \icomp \eta } ,
\end{equation}
\end{subequations}
where Coulomb integrals in the chemists' notation were introduced.

The nonequilibrium Green's function formalism provides a fast way to derive these equations. It can be illustrated by the computation of the lesser SE according to the diagrams in Fig.~\ref{diag:pt2}, where state indices on fermionic lines and frequencies associated with some propagators are indicated. For the ring diagram, standard diagrammatic rules 
(App.~\ref{app:D}) yield
\[
\Sigma^{\mathrm{1-ring},\,<}_{pq}(\w)
=\sum_{ibk}  2 (pi|bk) (qi|bk)\int\frac{d\nu}{2\pi}\frac{d\nu'}{2\pi} 
G^<_i(\w-\nu)G^<_k(\nu')G^>_b(\nu'-\nu).
\] 
Using definitions~\eqref{def:G:lessgtr}, and performing the two frequency integrals we obtain 
\[
\Sigma^{\mathrm{1-ring},\,<}_{pq}(\w)
=2\pi \icomp \sum_{ibk}  2 (pi|bk) (qi|bk)\delta(\omega -\epsilon_i  + \epsilon_b -\epsilon_k)
=2\icomp \im{\Sigma_{pq}^{\mathrm{1-ring},\, ovo}(\omega)}.
\] 
For the second-order exchange, the diagrammatic rules yield 
\[
\Sigma^{\mathrm{SOX},\,<}_{pq}(\w)
=-\sum_{ibk} (pi|bk)(qk|bi)\int\frac{d\nu}{2\pi}\frac{d\nu'}{2\pi} 
G^<_i(\w-\nu')G^<_k(\w-\nu)G^>_b(\w-\nu-\nu').
\] 
Again, by performing the two frequency integrals, we obtain 
\[\Sigma^{\mathrm{SOX},\,<}_{pq}(\w)
=-2\pi \icomp\sum_{ibk} (pi|bk)(qk|bi)\delta(\omega -\epsilon_i  + \epsilon_b -\epsilon_k)
=2\icomp \im{\Sigma_{pq}^{\mathrm{SOX},\, ovo}(\omega)}.
\]
For unoccupied states, we verify that 
$\Sigma^{\mathrm{PT2},\,>}_{pq}(\w)=-2\icomp \im{\Sigma_{pq}^{\mathrm{PT2},\, ovo}(\omega)}$,  as expected.

We note in passing that there is a correspondence between the nature of intermediate electronic states (occupied or virtual) in the diagrams in Fig.~\ref{diag:pt2} and the partition type (determined by the distribution $+$/$-$ over the internal vertices). In this example, this is a one-to-one correspondence because here only the lesser ($G^<\equiv G^{-+}$) and the greater ($G^>\equiv G^{+-}$) GFs are involved. Since the Coulomb interaction is instantaneous, no other partitions are possible. However, we will see examples involving time-ordered ($G^{--}$) or anti-time-ordered ($G^{++}$) GFs. These contain contributions from both occupied and virtual states, and thus the correspondence between partitions and the nature of intermediate states is no longer one-to-one. A useful rule of thumb is that partitions uniquely determine the scattering mechanisms, but not the character of the intermediate states.

The sum of the 1-ring and SOX terms is PSD because they are formed by gluing a single half-diagram with itself, without and with permutation of two fermionic lines, respectively (the argument is detailed in Ref.~\citenum{stefanucci_prb2014}). We remind the reader that the half-diagrams contain vertices on the same branch of the Keldysh contour, and they are obtained by dissecting the original diagrams along the dashed blue lines, Fig.~\ref{diag:pt2}. Independent of the diagrammatic argument, a direct proof of positive definiteness can be given. To this end, we manipulate the terms in Eqs.~\eqref{eq:1-ring:o} and \eqref{eq:soxovo} using a $i \leftrightarrow k$ symmetry and obtain

%
%
%

\begin{equation}
\label{eq:pt2_ovo2}
\Sigma_{pq}^{\mathrm{PT2},\, ovo}(\omega)
  = \sum_{ibk} 
   \frac{ (pi|bk) \cdot (qi|bk) + \frac{1}{2} (pi||bk) \cdot (qi||bk)}
        {\omega -\epsilon_i  + \epsilon_b -\epsilon_k   - \icomp \eta },
\end{equation}
where we introduced the antisymmetrized Coulomb matrix elements $(pi||bk)=(pi|bk) - (pk|bi)$. Now it is not difficult to verify that a projection of $\Sigma_{pq}^{\mathrm{PT2} \, ovo}(\omega)$ on an arbitrary state $x$ yields a PSD rate:
\[
- \icomp \langle x |\Sigma^{\mathrm{PT2},<}(\omega)|x\rangle=  2\im{\langle x |\Sigma^{\mathrm{PT2},\, ovo}(\omega)}|x\rangle=2\pi \sum_{ibk} 
\left(X_{ibk}^2 +  \tfrac{1}{2}Y_{ibk}^2\right)\delta(\omega -\epsilon_i  + \epsilon_b -\epsilon_k) \ge 0,
\]
with $X_{ibk}=\sum_p x_p(pi|bk)$ and $Y_{ibk}=\sum_p x_p(pi||bk)$. In the derivation, we used the Sokhotski–Plemelj formula 
\begin{equation}
 \label{eq:sokhotski}
 \lim_{\eta \to 0^+} \mathrm{Im} \frac{1}{\omega - E \pm \icomp \eta} = 
   \mp \pi \delta( \omega - E ) .
\end{equation}
Analogously, it can be shown that $ \icomp\langle x |\Sigma^{\mathrm{PT2},>}(\w)|x\rangle=-2\im{\langle x |\Sigma^{\mathrm{PT2} \, vov}(\omega)}|x\rangle\ge0$. This mathematical derivation confirms the numerical observation drawn from the Neon atom in Fig.~\ref{fig:neon1}. 

Note that the poles of 1-ring, SOX, and therefore PT2, are located 
at the bare energies $\epsilon_i - \epsilon_b + \epsilon_k$ 
or $\epsilon_a - \epsilon_j + \epsilon_c$ .
Since we will refer to these poles later, we assign names to them here: \hhp and \pph poles.

\subsection{$GW$ is PSD\label{sec:gw}}

Demonstrating that the one-shot $GW$ self-energy is PSD is straightforward 
once the analytic expressions of $G$ and $W_p$ are inserted in the
$GW$ convolution integral.
After the residue theorem is applied, the correlation part of the $GW$ self-energy is expressed as
the sum of the two following terms\cite{tiago_prb2006,vansetten_jctc2013,bruneval_cpc2016}:
\begin{subequations}
\label{eq:gw}
\begin{equation}
\label{eq:gwo}
\Sigma_{pq}^{GW,\,o}(\omega)
  = \sum_{ks} 
   \frac{ w_s^{pk} \cdot w_s^{qk} }
        {\omega -\epsilon_k  + \Omega_s  - \icomp \eta }
\end{equation}
and
\begin{equation}
\Sigma_{pq}^{GW,\,v}(\omega)
\label{eq:gwv}
  = \sum_{cs} 
   \frac{ w_s^{pc} \cdot w_s^{qc} }
        {\omega -\epsilon_c  - \Omega_s  + \icomp \eta }.
\end{equation}
\end{subequations}
It is instructive to derive these equations with NEGF formalism. By diagrammatic rules, we have
\begin{align*}
    \Sigma^{GW,\,<}_{pq}(\w)&=\icomp \sum_{ks} w_s^{pk}  w_s^{qk} \int\frac{d\nu}{2\pi}G^<_{k}(\w-\nu)W^<_s(\nu),
\end{align*}
where $W^<_s(\nu)=-2\pi\icomp\delta(\nu+\Omega_s)$ is defined by Eq.~\eqref{def:W:lessgtr}. Performing the frequency integral, we obtain
\begin{align}
     \Sigma^{GW,\,<}_{pq}(\w)&=2\pi \icomp \sum_{ks} w_s^{pk}  w_s^{qk} \delta(w-\epsilon_k  + \Omega_s )=2\icomp \im{\Sigma_{pq}^{GW,\,o}(\omega)}.
     \label{eq:Sgm:GW:lss}
\end{align}
Now, taking a projection on an arbitrary state $x$, we obtain a positive rate:
\begin{align*}
    -\icomp\langle x |\Sigma^{GW,\,<}(\omega)|x\rangle=  2\im{\langle x |\Sigma^{GW,\, o}(\omega)}|x\rangle=
    2\pi\sum_{ks} Z_{ks}^2\delta(w-\epsilon_k  + \Omega_s )\ge0,
\end{align*}
where $Z_{ks}=\sum_p x_p w_s^{pk}$. The contribution of virtual states ($\Sigma^{GW,>}$) can be treated similarly, and one shows that 
$ \icomp\langle x |\Sigma^{GW,\,>}(\w)|x\rangle=-2\im{\langle x |\Sigma^{GW,\, v}(\omega)}|x\rangle\ge0$. Thus, $GW$ is a PSD approximation. Furthermore, putting the explicit expression~\eqref{eq:Sgm:GW:lss} in correspondence with the diagram~\ref{diags:SE}(b), it becomes clear that its constituent half-diagrams are represented by the $w_s^{pk}$ and $w_s^{qk}$ matrix elements. In the next example, we will see that they also appear as a part of more complicated diagrams.

Note the specific location of the $GW$ poles, $\epsilon_k - \Omega_s$ and $\epsilon_c + \Omega_s$,
that include RPA excitation energies $\Omega_s$, instead of bare electron-hole excitations in the case of PT2 and SOX.

\subsection{$GW+G3W2$ is not PSD\label{sec:G3W2}}

Now turning to the $G3W2$ correction beyond $GW$, the analytic expression confirms the PSD-violation
we observed in Fig.~\ref{fig:neon1}.

Let us focus on a specific term in $G3W2$.
If this term is non-PSD and if it is not compensated by another term in the expression, this is sufficient
to show the non-PSD behavior of the overall approximation.
We then scrutinize the only term in $G3W2$ that involves three occupied states and, therefore, cannot be compensated [Eq.(13a) in Ref.\citenum{bruneval_jctc2024}]:
\begin{equation}
\label{eq:g3wp2ooo}
\Sigma_{pq}^{G3W_p2, \, ooo}(\omega)
=
\sum_{t s}
\sum_{i j k } 
\frac{w_t^{pi} \cdot w_t^{jk} \cdot w_s^{qk} \cdot w_s^{ij}}
{(\omega  - \epsilon_i + \Omega_t - \icomp \eta) \cdot (\omega - \epsilon_j + \Omega_t + \Omega_s  - \icomp \eta )
\cdot (\omega - \epsilon_k + \Omega_s  - \icomp \eta)}  .
\end{equation}
This expression involving three intermediate occupied states was obtained using the time-ordered formalism. Turning to the corresponding lesser SE component, further insights into the scattering mechanisms can be gained. There are three possible diagram partitions 1) $-+++$, 2) $--++$, and 3) $---+$, where we traverse the diagram's vertices from left to right (App.~\ref{app:D}). Note that the $-+-+$ partition, familiar from the analysis of the SOX diagram in Sec.~\ref{sec:pt2}, is excluded by the choice of intermediate states, and conversely, none of the present partitions are possible for the SOX diagram because they involve $W_p$ connecting different branches of the Keldysh contour (in contrast to $v$, $W_p$ is time non-local). Partitions 1 and 3 correspond to the hole scattering mechanism involving one neutral excitation (one fermionic and one bosonic line crossing the half-diagram border), whereas partition 2 corresponds to a higher-order process in which a hole scatters, emitting \emph{two} neutral excitations (one fermionic and bosonic lines crossing the border). The mechanisms can also be identified by inspecting the three terms in the denominator of Eq.~\eqref{eq:g3wp2ooo}.

Since 1 and 3 correspond to the same mechanism, they give rise to a \emph{double pole} when $i=k$ and $s=t$. This 
destroys the Lehmann decomposition of the self-energy \cite{winter_study_1972,cederbaum_jchemphys1975}, which is a fundamental property. The second problem becomes evident when we perform a partial fraction decomposition of Eq.~\eqref{eq:g3wp2ooo} and focus, for instance, on the $-\icomp\Sigma_{pq}^<=2\im{\Sigma_{pq}^{G3W_p2, \, ooo}}$ term proportional to $\delta(\omega - \epsilon_k + \Omega_s)$. Introducing a new matrix element
\begin{align}
  \widetilde{w}_s^{kp}&=\sum_{t}\sum_{i j}\frac{w_t^{pi}  \cdot w_s^{ij} \cdot w_t^{jk}} {(\e_k-\e_i -\W_s + \W_t) \cdot (\e_k - \e_j + \W_t)} \,,\label{eq:wt}
\end{align}
it can be written as 
\begin{align}
 -\icomp\Sigma_{pq}^<(\w)&\sim \widetilde{w}_s^{kp} w_s^{kq}\delta(\w-\e_k+\W_s).\label{eq:F1}
\end{align}
It is clear that this expression is not PSD, as $\widetilde{w}_s^{kp}$ contains a product of terms without a definite sign. Diagrammatically, $w_s^{kq}$ corresponds to a half-diagram with a single interaction vertex (which already appeared in the analysis of $GW$ self-energy in Sec.~\ref{sec:gw}), whereas $\widetilde{w}_s^{kp}$ is built of three interaction vertices. To ensure the PSD property, one would need, in addition to cross-terms $\widetilde{w}_s^{kp} w_s^{kq}$, to include the products $w_s^{kp}w_s^{kq}$ and $\widetilde{w}_s^{kp}\widetilde{w}_s^{kq}$.

\section{Design of a PSD-compliant 2SOSEX self-energy}
\label{sec:theory}

\begin{figure*}
\includegraphics[width=0.95\columnwidth]{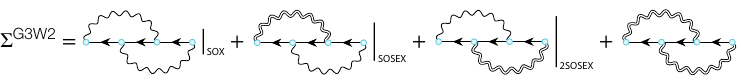} 
\caption{
Diagrammatic representation of the four terms contributing to the $G3W2$ approximation. Thin way lines stand for bare Coulomb interaction $v$, whereas  double wavy lines---for the frequency-dependent part $W_p(\w)$ of the screened interaction $W(\w)$. We therefore have $W(\w)=v+W_p(\w)$.
}
\label{diag:g3w2}
\end{figure*}
We just saw in the previous section that the $G3W2$ correction to $GW$ induces undesirable non-PSD terms in the self-energy. Moreover, some of them contain double poles. Leaving the multiple-pole issues aside\cite{pavlyukh_pade_2017}, we focus here on the positivity constraint.
In a recent series of works,\cite{stefanucci_prb2014,uimonen_prb2015,pavlyukh_prl2016} it was 
shown how the inclusion of higher-order terms could restore the PSD behavior of any approximation to the self-energy.
We will focus here on the 2SOSEX approximation to $G3W2$, which is free of double poles and therefore easier to handle mathematically. Moreover, in previous work it was shown\cite{bruneval_jctc2024} that 2SOSEX avoids many of the numerical difficulties of $G3W2$ while giving similar results. The proof will be made in two main steps:
1. all the \pph and \hhp poles in $G3W2$ compensate;
2. all the other poles in 2SOSEX can be complemented so to yield a PSD expression.

Let us recap the SOSEX and 2SOSEX approximations.
It is customary to split the static and dynamic interactions following Eq.~(\ref{eq:wp}).
Introducing this decomposition in $G3W2$ gives rise to 4 terms:
\begin{equation} 
\label{eq:g3w2}
 \Sigma^{G3W2}
    = 
     \underbrace{
       \underbrace{
         \underbrace{ \icomp^2 G v G v G }_\mathrm{SOX}
             +  \icomp^2  G v G W_p G}_\mathrm{SOSEX}
               +  \icomp^2 G W_p G v G }_\mathrm{2SOSEX}
                    +  \icomp^2  G W_p G W_p G  ,
\end{equation}
where the indices have been dropped for simplicity.
The first term in Eq.~(\ref{eq:g3w2}) is nothing else but the SOX self-energy of Eqs.~(\ref{eq:soxovo}-\ref{eq:soxvov}).
The first two terms form the SOSEX approximation of Ren and coworkers \cite{ren_prb2015}.
The three first terms were named 2SOSEX in our previous work \cite{bruneval_jctc2024}
and finally, the four terms are the fully dynamic $G3W2$ expression.

The argument in favor of introducing 2SOSEX is the observation that there is no reason to select the first term $G v G W_p G$
and not the second one $G W_p G v G$. As a matter of fact, for diagonal expectation values, they are equal.

In the following, we will introduce the analytic expression of SOSEX and show how the \pph and \hhp poles compensate.

\subsection{Analytic expression of the SOSEX self-energies}

Let us begin with the introduction of the SOSEX term in the self-energy.
We will use the same strategy as in Ref.~\citenum{bruneval_jctc2024}.
The terms will be grouped by occupied or virtual molecular orbitals for the 3 Green's function in $G v G W_p G$.
The static $v$ imposes that the first two occupations differ.
Then, instead of the 8 contributions present in $G W_p G W_p G$, only 4 will survive.

The energy convolutions can be performed with the application of the residue theorem.
Symbolic calculators, such as the open-source Wolfram Engine, can be of help to obtain the following expressions:
\begin{subequations}
\label{eq:sosex}
\begin{align}
\label{eq:sosex:ovo}
\Sigma_{pq}^{\mathrm{SOSEX},\, ovo}(\omega)
  =& \sum_{s} \sum_{ibk} \frac{ (pi|bk) \cdot w^{qk}_s \cdot w^{ib}_s}
                             {\omega -\epsilon_i + \epsilon_b -\epsilon_k - \icomp \eta } 
  \left[
  \frac{1}{\epsilon_b -\epsilon_i +\Omega_s } 
  +\frac{1}{\omega -\epsilon_k + \Omega_s - \icomp \eta  }
 \right] \\
\label{eq:sosex:vov}
\Sigma_{pq}^{\mathrm{SOSEX},\, vov}(\omega) 
 =& \sum_{s} \sum_{a j c} \frac{ (pa|jc) \cdot w_s^{qc} \cdot w_s^{aj} } {\omega -\epsilon_a + \epsilon_j -\epsilon_c + \icomp \eta } 
\left[
\frac{1}{\epsilon_a -\epsilon_j+\Omega_s }
-\frac{1}{\omega -\epsilon_c -\Omega_s + \icomp \eta}
\right] \\
\label{eq:sosex:voo}
\Sigma_{pq}^{\mathrm{SOSEX},\, voo}(\omega) 
 =& \sum_{s} \sum_{a j k } 
  \frac{
 (p a| jk) \cdot w_s^{q k} \cdot w_s^{a j}
       }
       {\epsilon_a-\epsilon_j+\Omega_s }
 \cdot
  \frac{1}{\omega -\epsilon_k +\Omega_s  - \icomp \eta } \\
\label{eq:sosex:ovv}
\Sigma_{pq}^{\mathrm{SOSEX},\, ovv}(\omega) 
  =& \sum_{s} \sum_{ibc} 
     \frac{
       (pi|bc) \cdot  w^{qc}_s \cdot w^{ib}_s 
          }{\epsilon_b -\epsilon_i +\Omega_s }
    \cdot
     \frac{1}{\omega - \epsilon_c -\Omega_s  + \icomp \eta }  .
\end{align}
\end{subequations}
Though these equations were already used in our previous work \cite{bruneval_jctc2024}, the analytic expressions were not written out explicitly. To form 2SOSEX, one should consider the alternate term $G W_p G v G$ whose
analytic expression is the same as SOSEX in Eqs.~(\ref{eq:sosex:ovo}-\ref{eq:sosex:ovv}) but with swapped indices [$(p, i, a, k, c) \leftrightarrow (q, k, c, i, a)$].


The expressions in Eq.~(\ref{eq:sosex:ovo}) and Eq.~(\ref{eq:sosex:vov}) contain products of two poles. To disentangle these, one can apply partial fraction decomposition, and after some tedious manipulations or using a symbolic calculator, one reaches these factorized expressions:
\begin{subequations}
\begin{align}
\label{eq:SOSEXrovo}
\Sigma_{pq}^{\mathrm{SOSEX},\, ovo}(\omega)
  =& \sum_{s} \sum_{ibk} (pi|bk) \cdot w^{qk}_s \cdot w^{ib}_s \nonumber\\
   &\times \left[
    \frac{1}{\omega -\epsilon_i + \epsilon_b -\epsilon_k - \icomp \eta}
    \cdot
    \frac{-2 \Omega_s}
         { \left( \epsilon_b - \epsilon_i \right)^2 - \Omega_s^2 }
   \right.  \nonumber\\ & ~~~~~\left.
  + \frac{1}{ \omega -\epsilon_k +\Omega_s  - \icomp \eta }
  \cdot
    \frac{1}{ \epsilon_b -\epsilon_i -\Omega_s }
 \right]  \\
\Sigma_{pq}^{\mathrm{SOSEX},\, vov}(\omega)
  =& \sum_{s} \sum_{ajc} (pa|jc) \cdot w^{qc}_s \cdot w^{aj}_s \nonumber\\
   & \times \left[
    \frac{1}{\omega -\epsilon_a + \epsilon_j -\epsilon_c + \icomp \eta}
    \cdot
    \frac{-2 \Omega_s }
         { \left( \epsilon_a - \epsilon_j \right)^2 - \Omega_s^2}
\right.  \nonumber\\ & ~~~~~\left.
  + \frac{1}{ \omega -\epsilon_c -\Omega_s  + \icomp \eta }
    \cdot
    \frac{1}{ \epsilon_a -\epsilon_j -\Omega_s }
 \right] .
\end{align}
\end{subequations}

These expressions have the advantage to show that contributions with \hhp and \pph poles can be treated separately from those with RPA excitation energies.
And indeed, we will show in the next subsection that the contributions with \hhp and \pph poles vanish in the full $G3W2$ self-energy.

\subsection{Compensation of bare energy \pph and \hhp poles}
\label{sec:compensation}

In order to show the cancellation of \hhp and \pph poles in $G3W2$, we will make use of an analytic formula that is not obvious at first sight.
The RPA elements $w_s$ and $\Omega_s$ fulfill the following equality:
\begin{equation}
\label{eq:relation}
 \sum_s w_s^{m n} w_s^{i a} \frac{- 2 ~\Omega_s}{ (\epsilon_a - \epsilon_i)^2 - \Omega_s^2 } = (m n | i a)  ,
\end{equation}
for any molecular orbital indices $(m,n)$ and any occupied-empty pair of MOs $(i, a)$.
This mathematical relation is demonstrated in Appendix~\ref{app:formula}.
In words, this relation expresses that the poles of the non-interacting susceptibility $\chi_0$ correspond to zeroes of the interacting RPA susceptibility $\chi$.

With this relation, we identify that the first term $\Sigma_{pq}^{\mathrm{SOSEX},\, ovo}(\omega)$ in Eq.~(\ref{eq:SOSEXrovo}),
is precisely the opposite of the SOX term,  $-\Sigma_{pq}^{\mathrm{SOX},\, ovo}(\omega)$ in Eq.~(\ref{eq:soxovo}).
This can be written as
\begin{subequations}
\label{id:SOSEX}
\begin{equation}
\label{eq:SOSEXrovo2}
 \Sigma_{pq}^{\mathrm{SOSEX},\, ovo}(\omega)
  =  
-\Sigma_{pq}^{\mathrm{SOX},\, ovo}(\omega) 
  + \sum_{s} \sum_{ibk} 
   \frac{(pi|bk) \cdot w^{qk}_s \cdot w^{ib}_s }
        { \epsilon_b -\epsilon_i -\Omega_s }
  \cdot
   \frac{1}{ \omega -\epsilon_k +\Omega_s  - \icomp \eta } .
\end{equation}

Symmetrically, we observe the analogous relation between the first term in $\Sigma_{pq}^{\mathrm{SOSEX},\, vov}(\omega)$ 
and  $-\Sigma_{pq}^{\mathrm{SOX},\, vov}(\omega)$ in Eq.~(\ref{eq:soxvov}).
\begin{equation}
\label{eq:SOSEXrvov2}
 \Sigma_{pq}^{\mathrm{SOSEX},\, vov}(\omega)
  = 
  -\Sigma_{pq}^{\mathrm{SOX},\, vov}(\omega) 
 + \sum_{s} \sum_{ajc} 
  \frac{ (pa|jc) \cdot w^{qc}_s \cdot w^{aj}_s }
       { \epsilon_a -\epsilon_j -\Omega_s }
 \cdot
   \frac{1}{ \omega -\epsilon_c -\Omega_s  + \icomp \eta } .
\end{equation}
\end{subequations}

This shows that in the SOSEX approximation, which is $\Sigma^\mathrm{SOX} + \Sigma^\mathrm{SOSEX}$, the SOX self-energy is completely canceled out.
In other words, in the SOSEX approximation, there is no pole located at bare energy differences.
The only poles that survive are located at $(\epsilon_a + \Omega_s)$ or $(\epsilon_i - \Omega_s)$, exactly those contained in standard $GW$.

Now, one can be worried about the 2SOSEX approximation we proposed in Ref.~\citenum{bruneval_jctc2024}.
Indeed, we can show following the exact same steps as above, that the left-screened self-energy SOSEX also contains
a $-\Sigma_{pq}^{\mathrm{SOX}}(\omega)$ contribution. 
This may look pathological at first sight.
Fortunately, we show now that this term is also cancelled out by a contribution in the dynamic $G3W_p2$ self-energy.

Starting from the analytic formulas for $G3W_p2$ in Ref.~\citenum{bruneval_jctc2024},
one can apply the partial fraction decomposition technique to separate the poles.
We focus on the term $\Sigma_{pq}^{G3W_p2\,ovo}(\omega)$ [Eq.~(13e) in Ref.~\citenum{bruneval_jctc2024}].
This permits us to isolate different kinds of poles, and in particular a \hhp pole:
\begin{multline}
\label{eq:g3wp2ovo}
\Sigma_{pq}^{G3W_p2,\,ovo}(\omega)
=
\sum_{t s}
\sum_{i b k} 
      w_t^{pi} \cdot w_t^{bk} \cdot w_s^{qk} \cdot w_s^{ib} \\
\times 
  \left[     
  \frac{1}{\omega  - \epsilon_i + \epsilon_b - \epsilon_k - \icomp \eta}
  \cdot
  \frac{-4 ~\Omega_s \cdot \Omega_t}{
        \left[ (\epsilon_b - \epsilon_i)^2 - \Omega_s^2\right]
        \cdot
        \left[ (\epsilon_b - \epsilon_k)^2 - \Omega_t^2\right]}
  \right. \\ \left.
    -
   \frac{1}{\omega - \epsilon_i + \Omega_t  - \icomp \eta }
   \cdot
    \frac{\epsilon_b -\epsilon_k +2 \Omega_s-\Omega_t}
  {\left(\epsilon_b - \epsilon_i + \Omega_s \right)
   \left(\epsilon_b - \epsilon_k - \Omega_t  \right)
   \left(\epsilon_i-\epsilon_k+\Omega_s-\Omega_t\right)}
  \right. \\ \left.
  +
   \frac{1}{\omega - \epsilon_k + \Omega_s - \icomp \eta }
   \cdot
   \frac{\epsilon_b -\epsilon_i +2 \Omega_t -\Omega_s}
        {\left( \epsilon_b - \epsilon_i - \Omega_s \right)
         \left( \epsilon_b - \epsilon_k + \Omega_t \right)
         \left(\epsilon_i-\epsilon_k+\Omega_s-\Omega_t\right)}
  \right. \\ \left.
  +
  \frac{1}{\omega -\epsilon_b - \Omega_s - \Omega_t + \icomp \eta }
  \cdot
  \frac{1}{\left( \epsilon_b  - \epsilon_i + \Omega_s \right)
  \left( \epsilon_b - \epsilon_k + \Omega_t  \right)}
  \right] .
\end{multline}

This equation contains the only \hhp pole in the complete $G3W_p2$ term.
Let us single out this contribution and discard the rest.
Applying twice the relation in Eq.~(\ref{eq:relation}), one can reach the desired result:
\begin{align}
\sum_{t s}
\sum_{i b k}
      w_t^{pi} \cdot w_t^{bk} \cdot w_s^{qk} \cdot w_s^{ib} 
  \left[
  \frac{1}{\omega  - \epsilon_i + \epsilon_b - \epsilon_k - \icomp \eta}
  \cdot
  \frac{-4 ~\Omega_s \cdot \Omega_t}{
        \left[ (\epsilon_b - \epsilon_i)^2 - \Omega_s^2\right] 
        \cdot
        \left[ (\epsilon_b - \epsilon_k)^2 - \Omega_t^2\right]} 
  \right] \nonumber \\
 = 
 - \sum_{i b k}
  \frac{1}{\omega  - \epsilon_i + \epsilon_b - \epsilon_k - \icomp \eta}
  \cdot
  \left[
\sum_t w_t^{pi} \cdot w_t^{bk}
  \frac{-2 ~\Omega_t}{
        (\epsilon_b - \epsilon_k)^2 - \Omega_t^2}
\right] \nonumber \\
 \cdot
\left[
\sum_s w_s^{qk} \cdot w_s^{ib}
  \frac{-2 ~\Omega_s}{ (\epsilon_b - \epsilon_i)^2 - \Omega_s^2 }
  \right] \nonumber \\
  = 
\Sigma_{pq}^{\mathrm{SOX}\,ovo}(\omega)  .
\end{align}

Of course, the same manipulations can be done for the particle scattering part, yielding the $\Sigma^{\mathrm{SOX},\,vov}$ term.
In total, this proves that the $G3W_p2$ contains a SOX contribution.

This derivation teaches us that the 2SOSEX approximation proposed in our previous work \cite{bruneval_jctc2024} retains
some \hhp and \pph contributions, but G3W2 would have none.
It would then make sense to construct a semi-static approximation to G3W2 that also incorporates the SOX contribution in the dynamic part $G3W_p2$.
That is why we introduce now an ``augmented'' version of 2SOSEX that we name ``2SOSEX-aug'' and that is defined as
\begin{equation}
\label{eq:aug:2SOSEX}
    \Sigma^{\mathrm{2SOSEX-aug}} = 
    2 \, \Sigma^\mathrm{SOX} 
         + \icomp^2 G v G W_p G
         +  \icomp^2 G W_p G v G  .
\end{equation}
This expression differs from 2SOSEX only by the factor 2 scaling the term $\Sigma^\mathrm{SOX}$. So far, the derivation was purely algebraic, and it seems hopeless to associate a diagrammatic form with it. However, the diagrammatic analysis is simple. 

\begin{figure*}[t]
\includegraphics[scale=1.2]{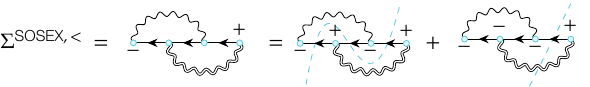} 
\caption{
Diagrammatic representation of the SOSEX self-energy.
}
\label{diag:SOSEX}
\end{figure*}

Let us return to the explicit SOSEX expressions~\eqref{eq:sosex}. To analyze hole scattering mechanisms, it is natural to look at the rate function associated with the lesser self-energy $-i\Sigma^<(\w)$. Taking the imaginary part of Eqs.~(\ref{eq:sosex:ovo} and \ref{eq:sosex:voo}), we see that there are two scattering mechanisms at play: involving RPA neutral excitations giving rise to $\delta\bigl(\w-(\epsilon_k-\W_s)\bigr)$, and involving bare particle-hole excitations giving rise to the $\delta\bigl(\w-(\epsilon_k-(\epsilon_b-\epsilon_i))\bigr)$ frequency dependence. The particle scattering processes are classified analogously by analyzing $i\Sigma^>(\w)$. 

\begin{figure*}[t]
\includegraphics[scale=1.2]{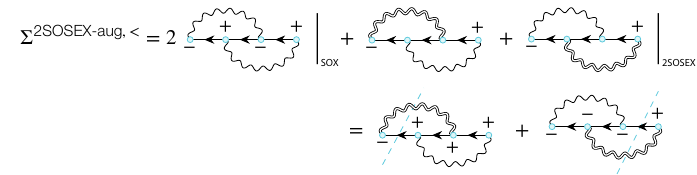} 
\caption{
Diagrammatic representation of Eq.~\eqref{eq:aug:2SOSEX}. The transformation is performed with the help of Eqs.~\eqref{id:SOSEX}, which originate from the identity~\eqref{eq:relation}.
}
\label{diag:aug:2SOSEX}
\end{figure*}

Just as the time-ordered SE, $\Sigma^{\lessgtr}$ can be analyzed diagrammatically. For $\Sigma^<$ we have to remember that outer vertices are situated on the $-$/$+$ branches, respectively, whereas internal vertices can take position at either Keldysh branch ($-$ or $+$), provided that i) there should be no isolated islands of certain sign, ii) bare Coulomb interaction is time-instantaneous, and therefore can only connect vertices on the same branch. Given these constraints, it turns out that the lesser SOSEX self-energy consists of just two partitions as depicted in Fig.~\ref{diag:SOSEX}.
It is not difficult to verify that they have distinct functional forms---they describe different physical mechanisms. The first partition involves 2 lesser and 1 greater GFs. Integrating over the intermediate times yields the  $\delta\bigl(\w-(\epsilon_k-(\epsilon_b-\epsilon_i))\bigr)$ dependence. Moreover, as was shown in Sec.~\ref{sec:compensation}, this partition is exactly canceled by the SOX term. The second partition involves 1 lesser GF and 1 lesser $W_p$ line. Integrating over the intermediate times yields the $\delta\bigl(\w-(\epsilon_k-\W_s)\bigr)$ dependence. Summarizing, Eq.~\eqref{eq:aug:2SOSEX} can be depicted diagrammatically as in Fig.~\ref{diag:aug:2SOSEX}. It is then obvious, considering the constituent half-diagrams, that this expression cannot be written as a sum of squares. Thus, this approximation is not manifestly PSD. We will improve upon this approximation in the next section.

Half-diagrams in Fig.~\ref{diag:aug:2SOSEX} have a clear diagrammatic meaning, analytic expressions can be derived using standard rules. However, it is of interest to support these derivations by transforming the time-ordered expressions. Now that we have shown that the \hhp and \pph poles disappear in 2SOSEX-aug, we focus on a scattering mechanism involving RPA neutral excitations. For occupied states, we therefore inspect terms corresponding to the poles $1/(\omega-\epsilon_k+\Omega_s-\icomp\eta)$. They originate from the $GW$ term given in Eq.~(\ref{eq:gwo}), Eq.~(\ref{eq:sosex:voo}) and the non-SOX term in Eq.~(\ref{eq:SOSEXrovo2}), and the corresponding terms from the symmetric self-energy $GW_pGvG$. Relabeling the indices and using the symmetry of $w_s^{ia} = w_s^{ai}$, we finally arrive at a compact expression analogous to $GW$:
\begin{subequations}
    
\begin{equation}
   \Sigma_{pq}^{GW+\mathrm{2SOSEX-aug},o} 
  = \sum_{ks} 
   \frac{ w_s^{pk} \cdot  w_s^{qk} + \tilde w_s^{pk} \cdot w_s^{qk} + w_s^{pk} \cdot \tilde w_s^{qk}  }
        {\omega -\epsilon_k  + \Omega_s  - \icomp \eta },
        \label{eq:aug-2SOSEX:o}
\end{equation}
where
\begin{equation}
\tilde w_s^{pk}
 = \sum_{a j} w_s^{aj} \cdot \left[
   \frac{(p a| j k)}{ \epsilon_a - \epsilon_j + \Omega_s }
 + \frac{(pj|ak)}{ \epsilon_a -\epsilon_j -\Omega_s }
   \right]
\end{equation}
In agreement with diagrammatic form in Fig.~\ref{diag:aug:2SOSEX}, Eq.~\eqref{eq:aug-2SOSEX:o} contains three terms.

For virtual states the analysis is analogous, it starts from Eqs.~(\ref{eq:gwv}), (\ref{eq:SOSEXrvov2}), and (\ref{eq:SOSEXrvov2}) and their $GW_pGvG$ counterparts and yields the following expression:
\begin{equation}
   \Sigma_{pq}^{GW+\mathrm{2SOSEX-aug},v} 
  = \sum_{cs} 
   \frac{ w_s^{pc} \cdot  w_s^{qc} + \tilde w_s^{pc} \cdot w_s^{qc} + w_s^{pk} \cdot \tilde w_s^{qc}  }
        {\omega -\epsilon_c  - \Omega_s  + \icomp \eta },
        \label{eq:aug-2SOSEX:v}
\end{equation}
where
\begin{equation}
\tilde w_s^{pc}
 = \sum_{a j} w_s^{aj} \cdot \left[
   \frac{(p a| j c)}{ \epsilon_a - \epsilon_j - \Omega_s }
 + \frac{(pj|ac)}{ \epsilon_a -\epsilon_j + \Omega_s }
   \right] .
\end{equation}
\end{subequations}
$\tilde w_s^{pk}$, $\tilde w_s^{pc}$ have the meaning of the first-order (in $v$) correction to hole-neutral excitation, particle-neutral excitation scattering processes, respectively.  
\subsection{Final expression: \psd}
\begin{figure*}[t]
\includegraphics[width=\columnwidth]{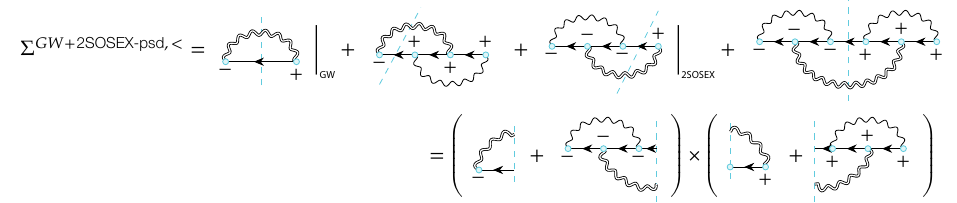} 
\caption{
Diagrammatic representation of the lesser SE component corresponding to Eq.~\eqref{eq:psd-2SOSEX:o}.
}
\label{diag:psd:2SOSEX}
\end{figure*}

Equations \eqref{eq:aug-2SOSEX:o} and \eqref{eq:aug-2SOSEX:v} are already very useful. They correspond directly to the diagrams in Fig.~\ref{diag:aug:2SOSEX}, and they are easy to evaluate numerically. Moreover, they clearly demonstrate that $\tilde w_s^{pk}$ and $\tilde w_s^{pc}$ are responsible for the overall non-PSD behavior. In fact, due to the energy differences in the denominators, these matrix elements do not have a definite sign. But adding extra terms quadratically proportional to $\tilde w$ allows us to complete a square in the numerators of Eqs.~\eqref{eq:aug-2SOSEX:o} and \eqref{eq:aug-2SOSEX:v} and yields manifestly PSD expressions:
\begin{subequations}
    \label{eq:psd:2SOSEX}
\begin{equation}
    \Sigma_{pq}^{GW+\mathrm{2SOSEX-psd},\,o }
  = \sum_{ks} 
   \frac{ ( w_s^{pk} + \tilde w_s^{pk} ) \cdot ( w_s^{qk} + \tilde w_s^{qk} )  }
        {\omega -\epsilon_k  + \Omega_s  - \icomp \eta }
        \label{eq:psd-2SOSEX:o}
\end{equation}

\begin{equation}
    \Sigma_{pq}^{GW+\mathrm{2SOSEX-psd},\,v} 
  = \sum_{cs} 
   \frac{ (w_s^{pc} + \tilde w_s^{pc}) \cdot (w_s^{qc} + \tilde w_s^{qc})  }
        {\omega -\epsilon_c  - \Omega_s  + \icomp \eta }
\end{equation}
\end{subequations}
A critical reader, however, would demand that correction terms should not be added \emph{ad hoc}, but rather have a clear diagrammatic meaning. Such interpretation, which involves, as expected, a partition of a certain third-order diagram, indeed exists and in depicted in Fig.~\ref{diag:psd:2SOSEX}. In this figure, we consider the lesser SE component describing scattering of occupied states and corresponding to Eq.~\eqref{eq:psd-2SOSEX:o}. The bottom row illustrates its complete square form.   

\begin{table}
\caption{Summary of the self-energy approximations considered in this work with their properties.
}
\label{tab:summary}
\begin{tabular}{lll}
\toprule
Approximation            & Free of \pph, \hhp poles & Positive semi-definite \\
\midrule
$GW$  & Yes & Yes \\
$GW$+SOX          & No  & No  \\
$GW$+SOSEX        & Yes & No  \\
$GW$+2SOSEX       & No  & No  \\
\aug              & Yes & No  \\
\psd              & Yes & Yes \\
$GW$+$G3W2$                  & Yes & No \\
\bottomrule
\end{tabular}
\end{table}

Let us conclude this section with Table~\ref{tab:summary} that recapitulates the analytic properties 
of the different approximations beyond $GW$ that we discussed in this work.
As shown above, standard $GW$ is devoid of \hhp and \pph poles and PSD and we were looking for
vertex corrections that are as mathematically sound as $GW$.
SOX and 2SOSEX corrections fail with respect to these two criteria.
\aug only fixes the no-\pph-\hhp pole property and finally \psd fixes both.
Going too far and adding all the contributions of $G3W2$ destroys again the PSD constraint.

\subsection{Connection to effective Hamiltonians}

Another way to look at PSD self-energies is through the lens of effective Hamiltonians of the general form 
\begin{equation}
\label{Heff}
    \mathbf{H} = 
    \begin{pmatrix}
        \mathbf{F} &\mathbf{C} \\
        \mathbf{C}^{\dagger} & \mathbf{D}
    \end{pmatrix} \,.
\end{equation}
Here, $\mathbf{F}$ is the Fock or generalized Kohn--Sham Hamiltonian with matrix elements $\epsilon_p\delta_{pp'}$, and $\mathbf{D}$ is a diagonal matrix with the matrix elements $D_{qs,q's'} = (\epsilon_q + \text{sgn}(\epsilon_q-\mu)\Omega_s)\delta_{qq'}\delta_{ss'}$, describing auxiliary states to which the single-particle states $\epsilon_p$ are coupled through $\mathbf{C}$ with matrix elements $C_{p,qs}$. Equation~(\ref{Heff}) can be downfolded onto the single-particle space, resulting in the self-energy
\begin{equation}
    \mathbf{F} + \mathbf{\Sigma}^c(\omega) = \mathbf{F} + \mathbf{C} \left[\omega\mathbf{1} - \mathbf{D}\right]^{-1} \mathbf{C}^{\dagger}\;.
\end{equation}
Since any matrix $\mathbf{B} = \mathbf{C}\mathbf{C}^{\dagger}$ is PSD, any self-energy which can be obtained through downfolding from an effective Hamiltonian is PSD.
Conversely, since any PSD self-energy admits a Lehmann representation,\cite{cederbaum_jchemphys1975} it can be represented as an effective single-particle Hamiltonian of the form in Eq.~(\ref{Heff}). Indeed, with $C_{p,qs} = w^{pq}_s$, the $GW$ self-energy from Eq.~(\ref{eq:gw})\cite{Bintrim2021, Monino2022, Tolle2023} is obtained, and setting $C_{p,qs} = w_s^{pq} + \tilde w_s^{pq}$, the \psd self-energy in Eq.~(\ref{eq:psd:2SOSEX}) is recovered. Replacing the RPA neutral excitation energies with the bare particle-hole pairs and modifying the couplings accordingly, one obtains the Hamiltonian corresponding to the PT2 self-energy.\cite{Backhouse2020} On the other hand, none of the non-PSD self-energy expressions presented in this work correspond to an effective Hamiltonian of the form of Eq.~(\ref{Heff}).
Though not numerically competitive, this alternate form is worth mentioning since it could help deriving exact properties in the future.

\section{Numerical results for molecular systems}
\label{sec:bench}

The new self-energies we have introduced in the previous section, namely 
\aug and \psd, have appealing formal properties.
However, the true test is the actual accuracy of these approximations
on real electronic systems.
In the following, we focus on molecular systems for valence and core electron $qp$ energies. All results shown are obtained with MOLGW.\cite{bruneval_cpc2016}

In the course of this study, we also evaluated the complete spectral functions,
but we found that the satellite structures in the spectral functions were very sensitive to the technical details,
basis set, mean-field starting point.
In our opinion, they are not conclusive pieces of information.

\subsection{Imaginary part of the self-energy of molecular examples}

\begin{figure}[hbt!]
\includegraphics[width=0.32\columnwidth]{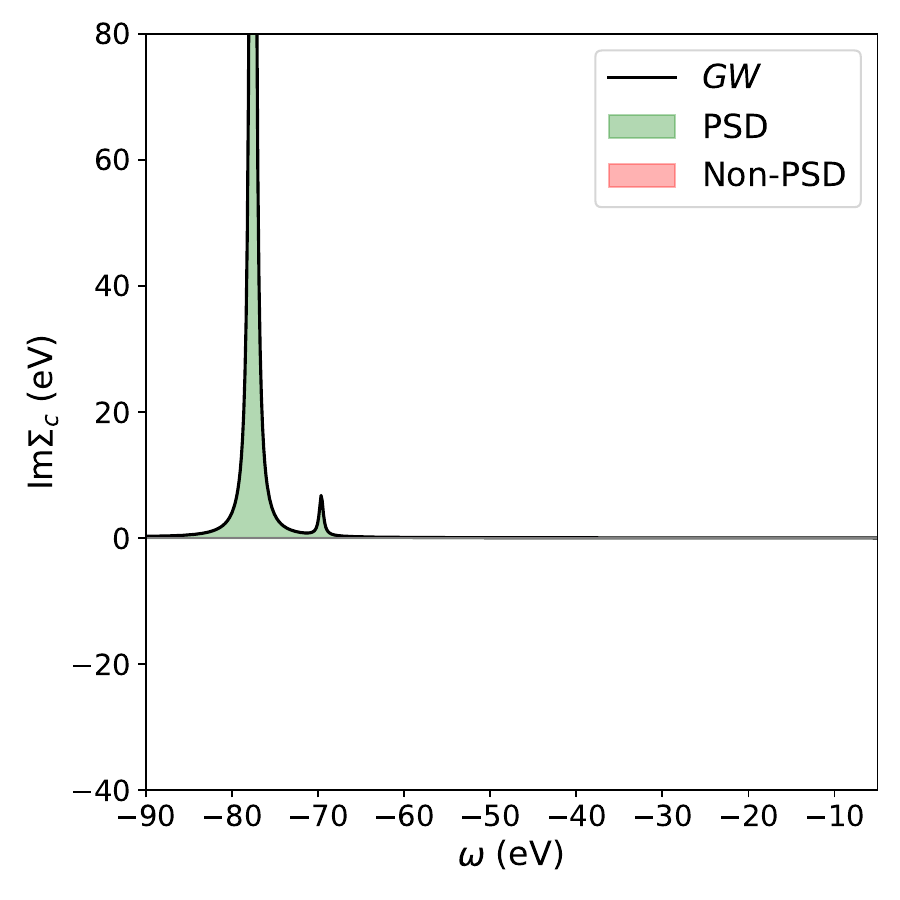}
\includegraphics[width=0.32\columnwidth]{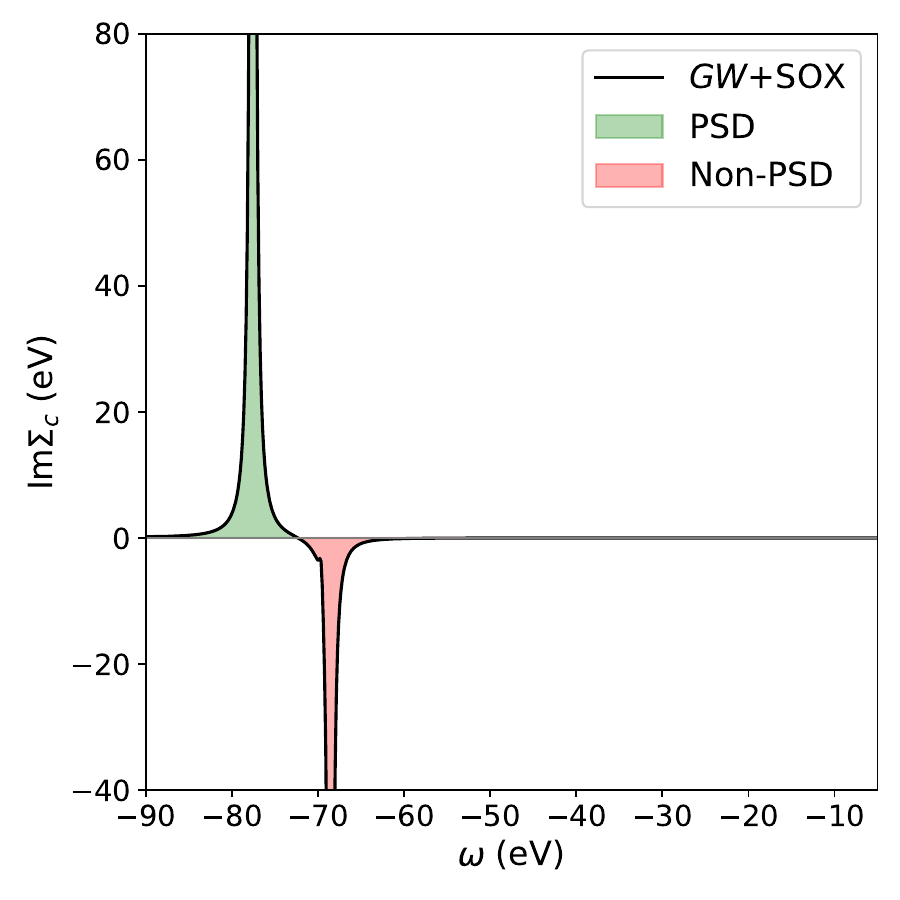}
\includegraphics[width=0.32\columnwidth]{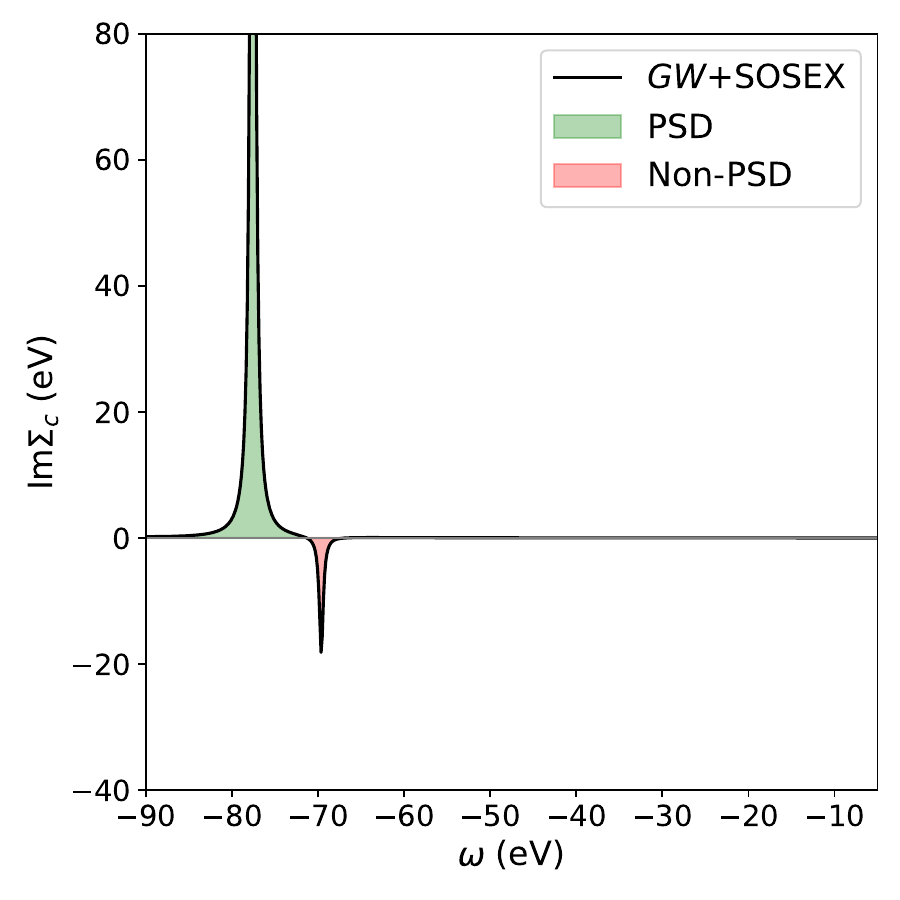} \\
\includegraphics[width=0.32\columnwidth]{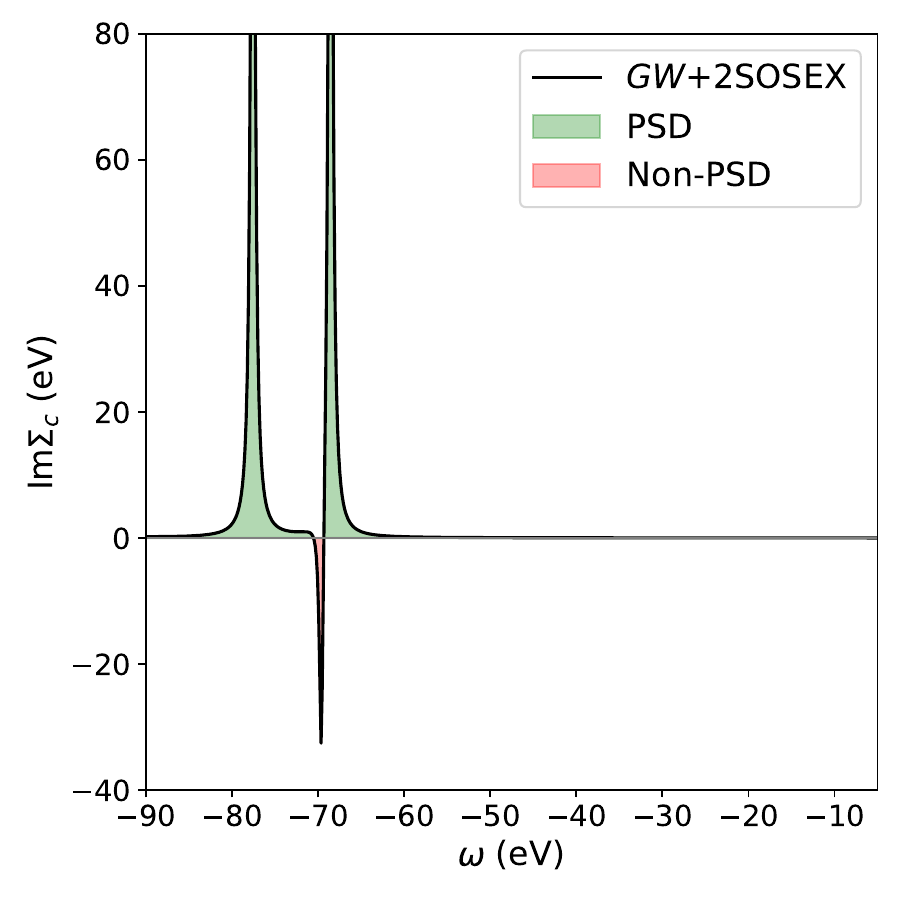}
\includegraphics[width=0.32\columnwidth]{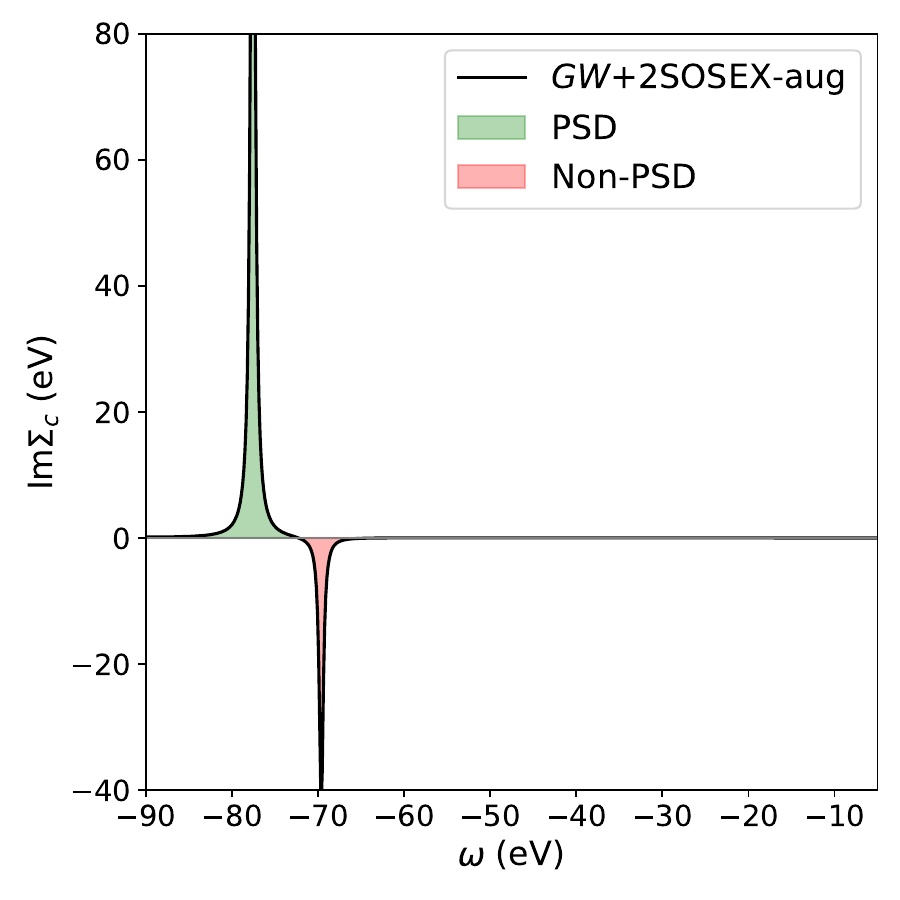}
\includegraphics[width=0.32\columnwidth]{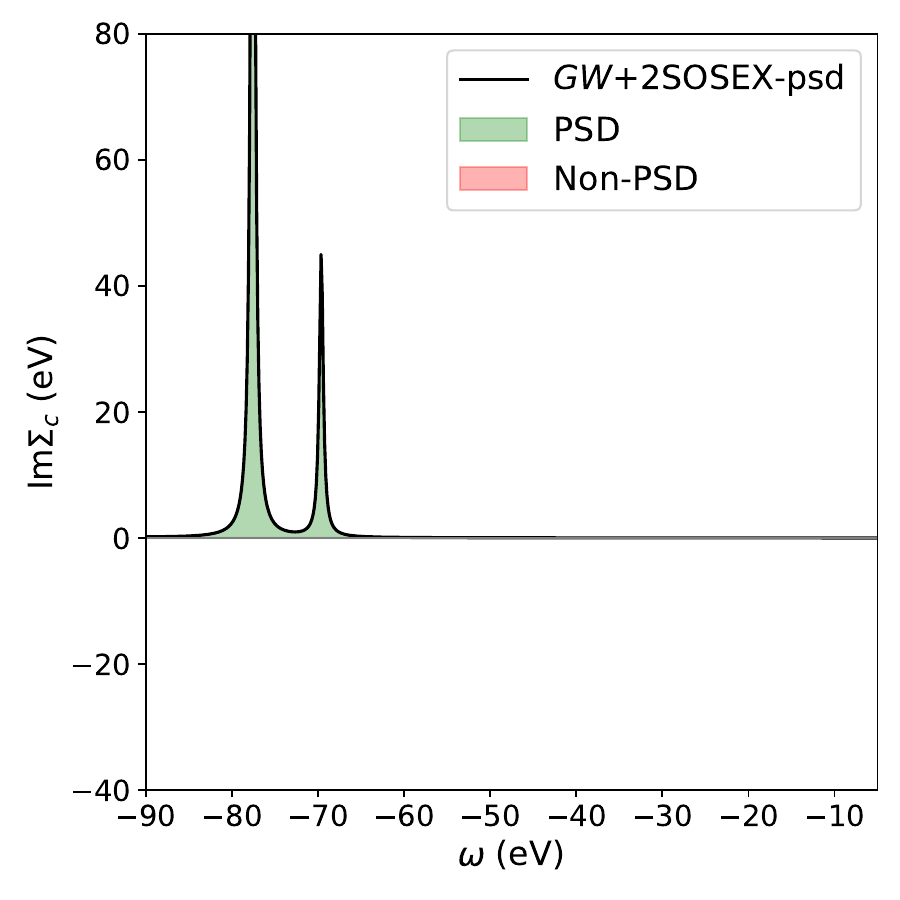}
\caption{
Imaginary part of time-ordered correlation self-energies for the Ne atom.
We plot the HOMO diagonal expectation value $\mathrm{Im} \langle \mathrm{HOMO} | \Sigma_c(\omega) | \mathrm{HOMO} \rangle$
for $GW$, $GW$+SOX, $GW$+SOSEX in the first row, and $GW$+2SOSEX, $GW$+2SOSEX-aug, $GW$+2SOSEX-psd in the second row.
Self-energies are evaluated as one-shot correction on top of HF.
}
\label{fig:neon2}
\end{figure}

Let us first illustrate the approximations gathered in Table~\ref{tab:summary} with
the Ne atom that we used in the introduction in Fig.~\ref{fig:neon1}.

Figure~\ref{fig:neon2} reports the imaginary part of 6 self-energy approximations starting from $GW$ and complementing it with vertex corrections.
$GW$+SOX is the worst of all with strong non-PSD weight. This confirms the known statement that $GW$+SOX is not producing accurate quasiparticle energies
\cite{marom_prb2012, bruneval_fchem2021}.
$GW$+SOSEX, which is devoid of \pph and \hhp poles, shows improved performance with respect to the PSD criterion, even though it is not perfect.
$GW$+2SOSEX has a strong positive contribution around -68~eV, which comes from the $-\Sigma^\mathrm{SOX}$ contribution.
Then in \aug, this contribution is eliminated. Remember that the only difference between $GW$+2SOSEX and \aug is the addition of an extra $\Sigma^\mathrm{SOX}$.

Finally, our \psd self-energy is perfectly PSD as expected. Its overall shape is much more similar to the original $GW$ than any one of the other approximations.
With the knowledge of the performance of $GW$ for molecules, it is a very positive sign that \psd stays rather similar to $GW$.

\begin{figure}[hbt!]
\includegraphics[width=0.32\columnwidth]{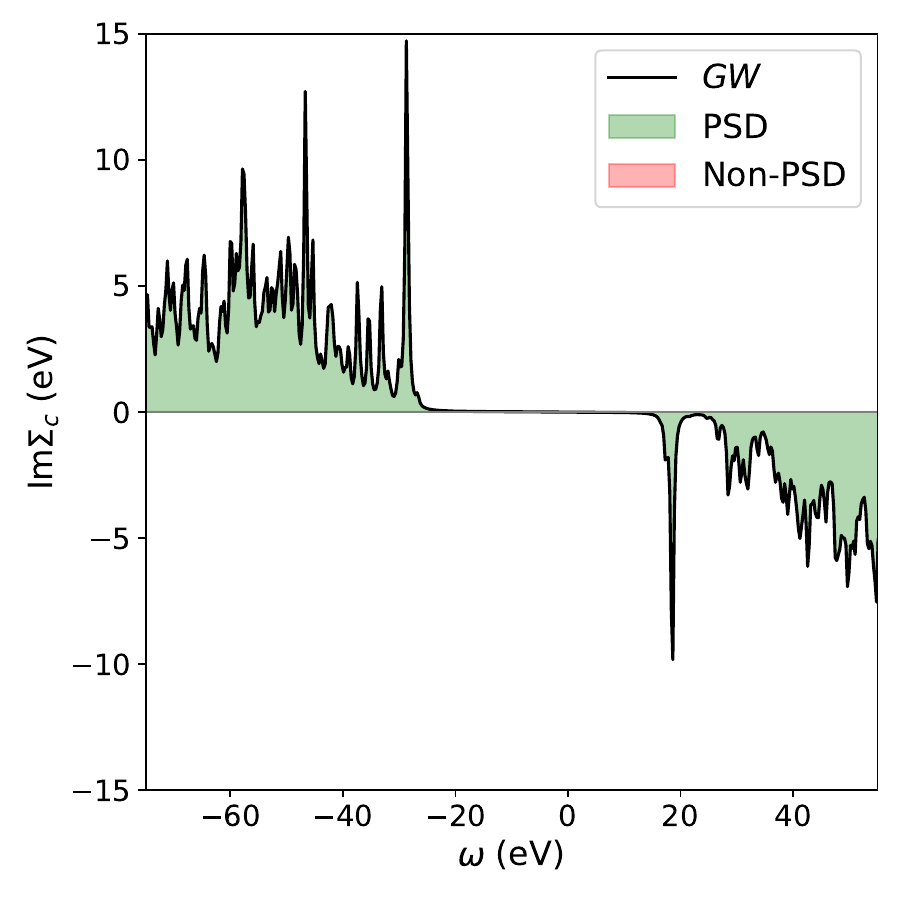}
\includegraphics[width=0.32\columnwidth]{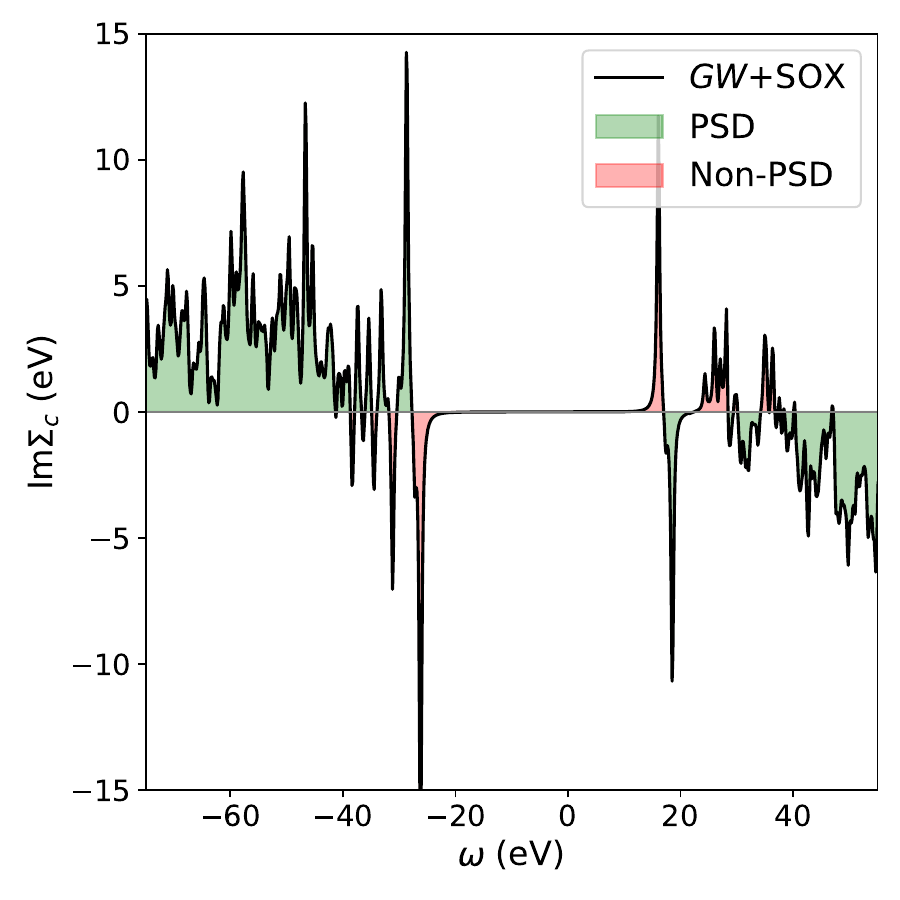}
\includegraphics[width=0.32\columnwidth]{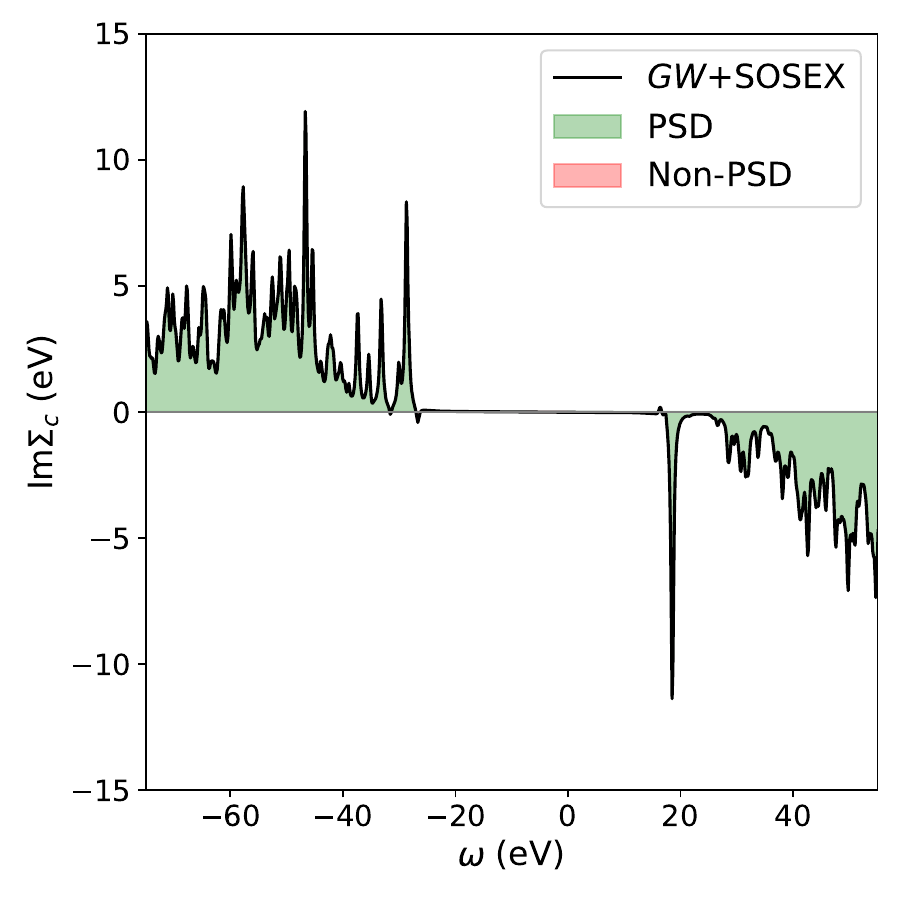} \\
\includegraphics[width=0.32\columnwidth]{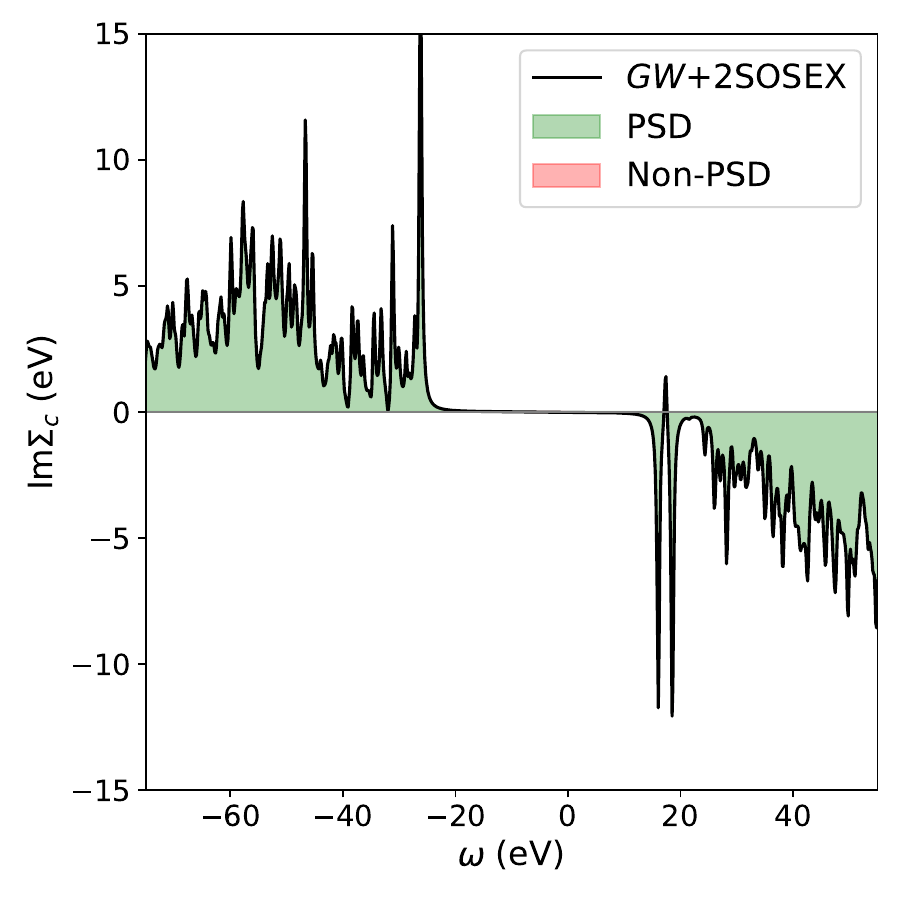}
\includegraphics[width=0.32\columnwidth]{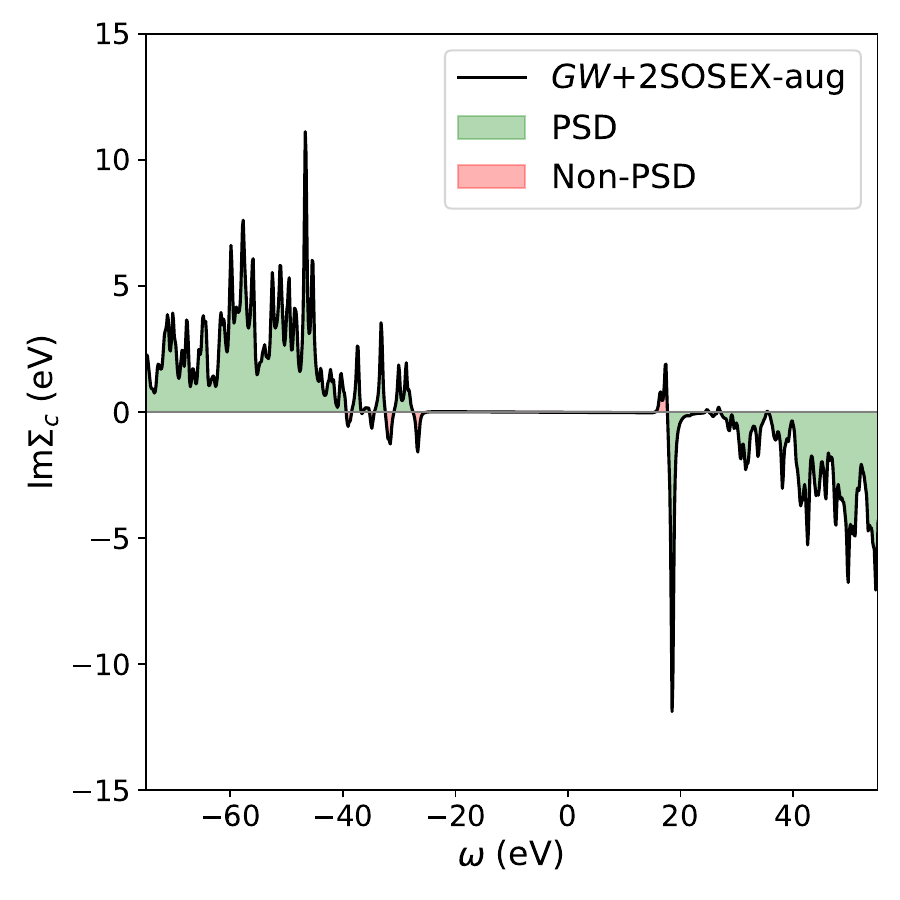}
\includegraphics[width=0.32\columnwidth]{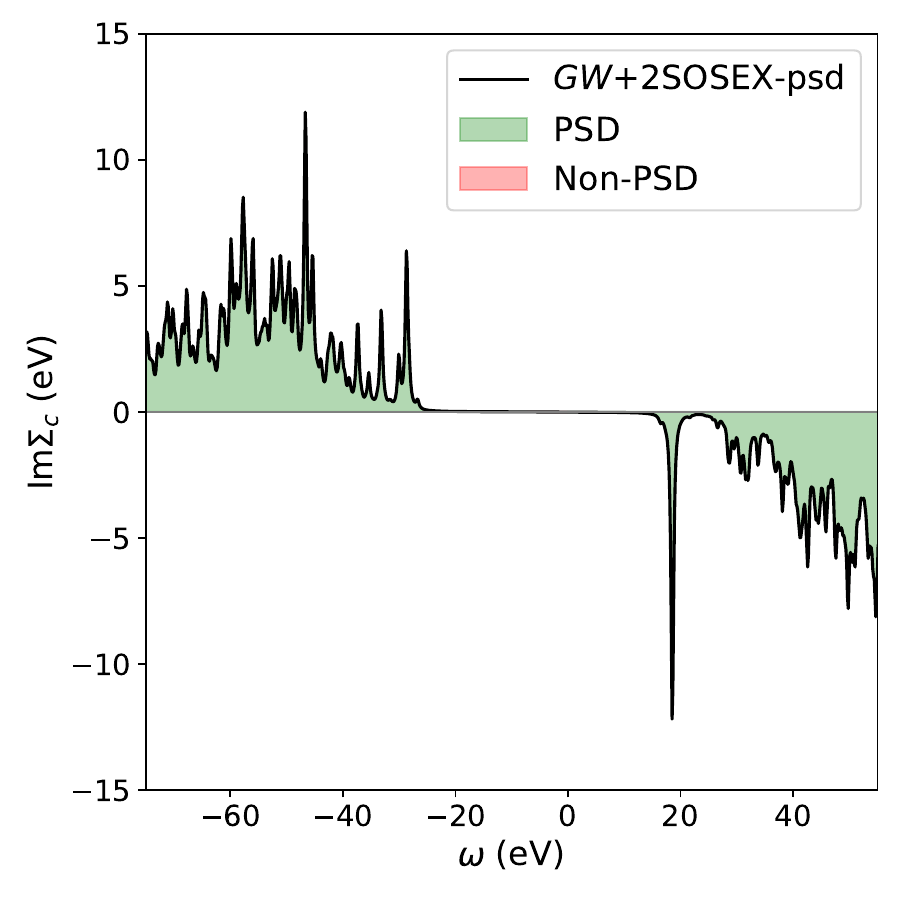}
\caption{
Imaginary part of time-ordered correlation self-energies for the benzene molecule, C$_6$H$_6$.
We plot the HOMO diagonal expectation value $\mathrm{Im} \langle \mathrm{HOMO} | \Sigma_c(\omega) | \mathrm{HOMO} \rangle$
for $GW$, $GW$+SOX, $GW$+SOSEX in the first row, and $GW$+2SOSEX, $GW$+2SOX+2SOSEX, $GW$+2SOX+2SOSEX-PSDized in the second row.
A broadening $\eta=0.01$~Ha was used.
}
\label{fig:benzene}
\end{figure}

Second, in Fig.~\ref{fig:benzene}, we show imaginary parts of self-energies for a more complicated example, benzene, where the poles almost form a continuum.
The conclusions we drew on the Neon case are confirmed, except that the non-PSD peaks are less pronounced.
Notice once again the similarity between $GW$ and \psd: the peaks are the same, just their intensity was modified.
This is consistent with the Eqs.~(\ref{eq:psd:2SOSEX}) that only differ from their $GW$ counterpart Eqs.~(\ref{eq:gw}) by the numerators.

%

\subsection{GW100 ionization potentials benchmark}

\begin{figure*}
\includegraphics[width=0.80\columnwidth]{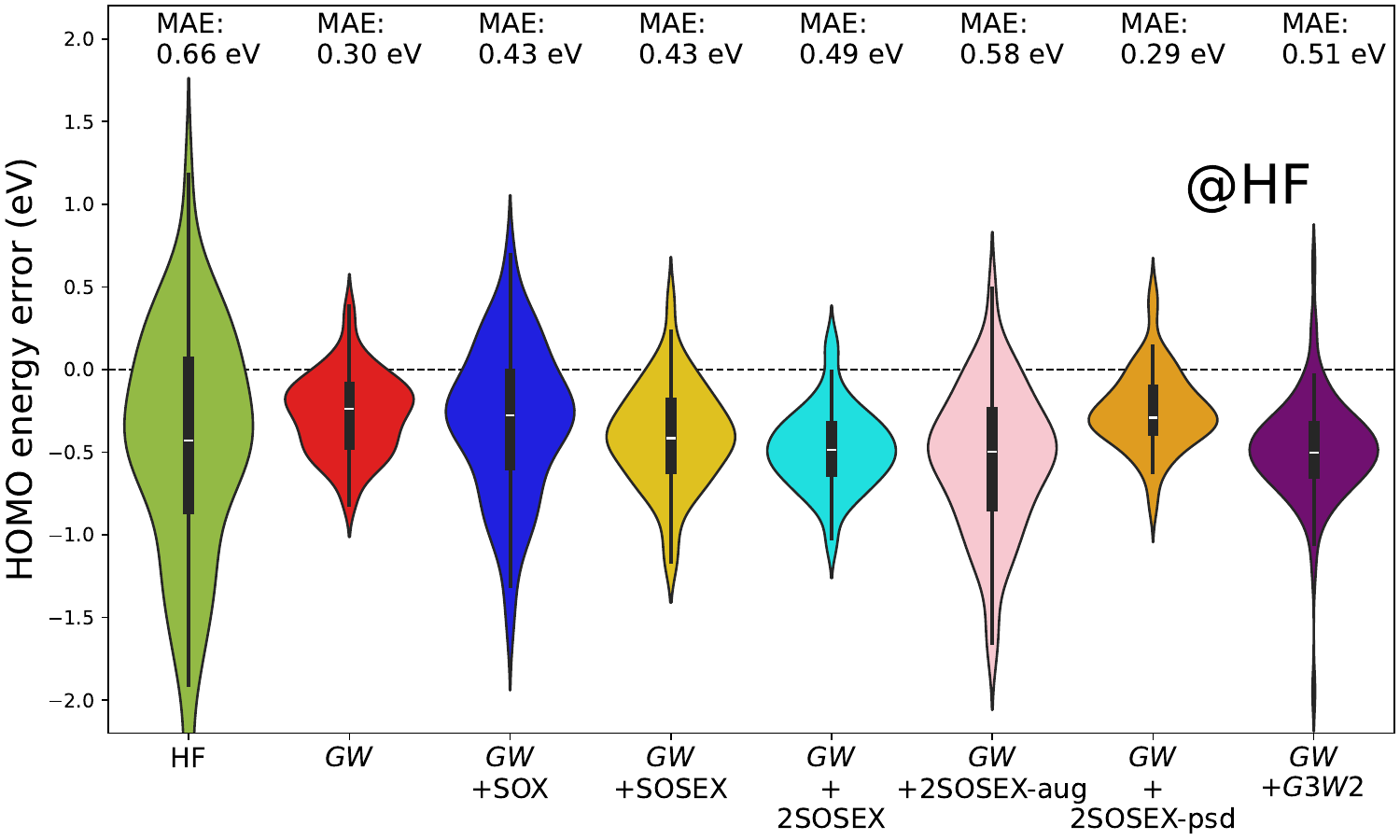}
\caption{
Error distribution of the HOMO energies with respect to a reference CCSD(T) $\Delta$SCF evaluation
for the 100 molecules contained in the GW100 set.
Different self-energies are evaluated as one-shot perturbations on top of HF.
}
\label{fig:gw100hf}
\end{figure*}

Thanks to the reference coupled-cluster ionization potentials from Refs.~\citenum{Krause2015} and~\citenum{bruneval_fchem2021}, 
the GW100 set \cite{vansetten_jctc2015} of 100 small- and medium-sized molecules can serve as a fair measure of the quality of a self-energy approximation for
HOMO quasiparticle energies.
In the following, we use the same Gaussian basis set as in the reference calculations, namely Def2-TZVPP, \cite{weigend_pccp2005} to mitigate the basis set effect on the conclusions.

Sticking to one-shot evaluation of the self-energy as assumed in our equations, there always exists a starting point dependence in the results
and this may obscure the conclusions.
Therefore, we proceed in two steps.
In a first step, we evaluate the self-energies using Hartree-Fock (HF) as a starting point, as is standard in quantum chemistry.
In a second step, we use improved starting points for selected self-energy approximations, as the $GW$ community often do.

Figure~\ref{fig:gw100hf} presents the GW100 HOMO energy (negative of the ionization potential) error distributions for 
all the approximations reported in Table~\ref{tab:summary}, plus the initial HF mean-field result.
HF is not a particularly good starting point for HOMO energies as we can see from the broad violin and the large mean absolute error (MAE).
$GW$ based on HF is rather good, with an MAE of 0.30~eV.
The subsequent additions to $GW$, namely $GW$+SOX, $GW$+SOSEX, $GW$+2SOSEX, are not as accurate as the plain $GW$ approximation.
Through the elimination of the bare energy \pph and \hhp poles, \aug choice performs noticeably worse than $GW$+2SOSEX.
This is quite unexpected.
Quite the opposite, \psd shows a reliability that is very comparable to $GW$ with an MAE of 0.29~eV and a narrow distribution.
For completeness, we also included the $GW$+$G3W2$ result from our previous work \cite{bruneval_jctc2024}.
This last approximation shows serious outliers.
Note that the analytic formula of \psd is way faster to evaluate than that of $GW$+$G3W2$ with a much improved accuracy.

\begin{figure*}
\includegraphics[width=0.45\columnwidth]{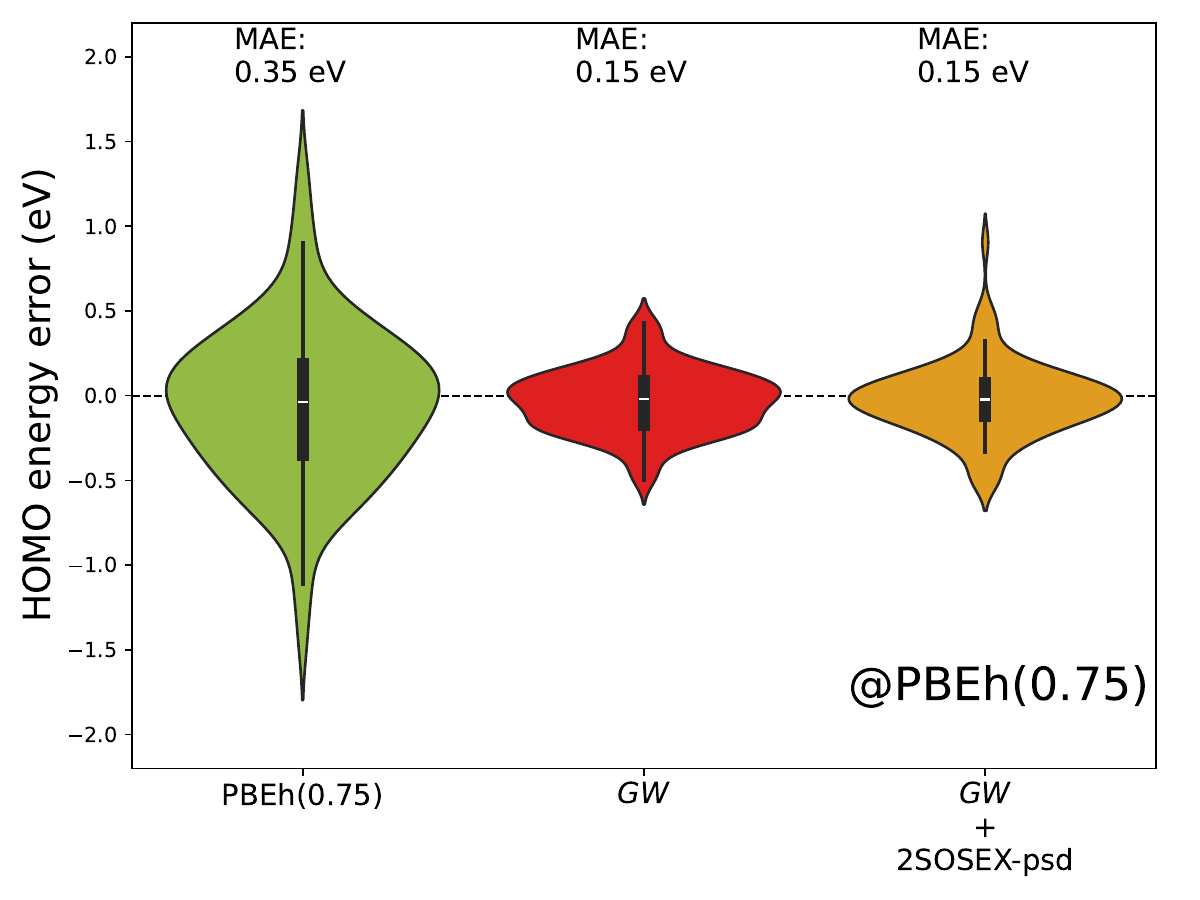}
\includegraphics[width=0.45\columnwidth]{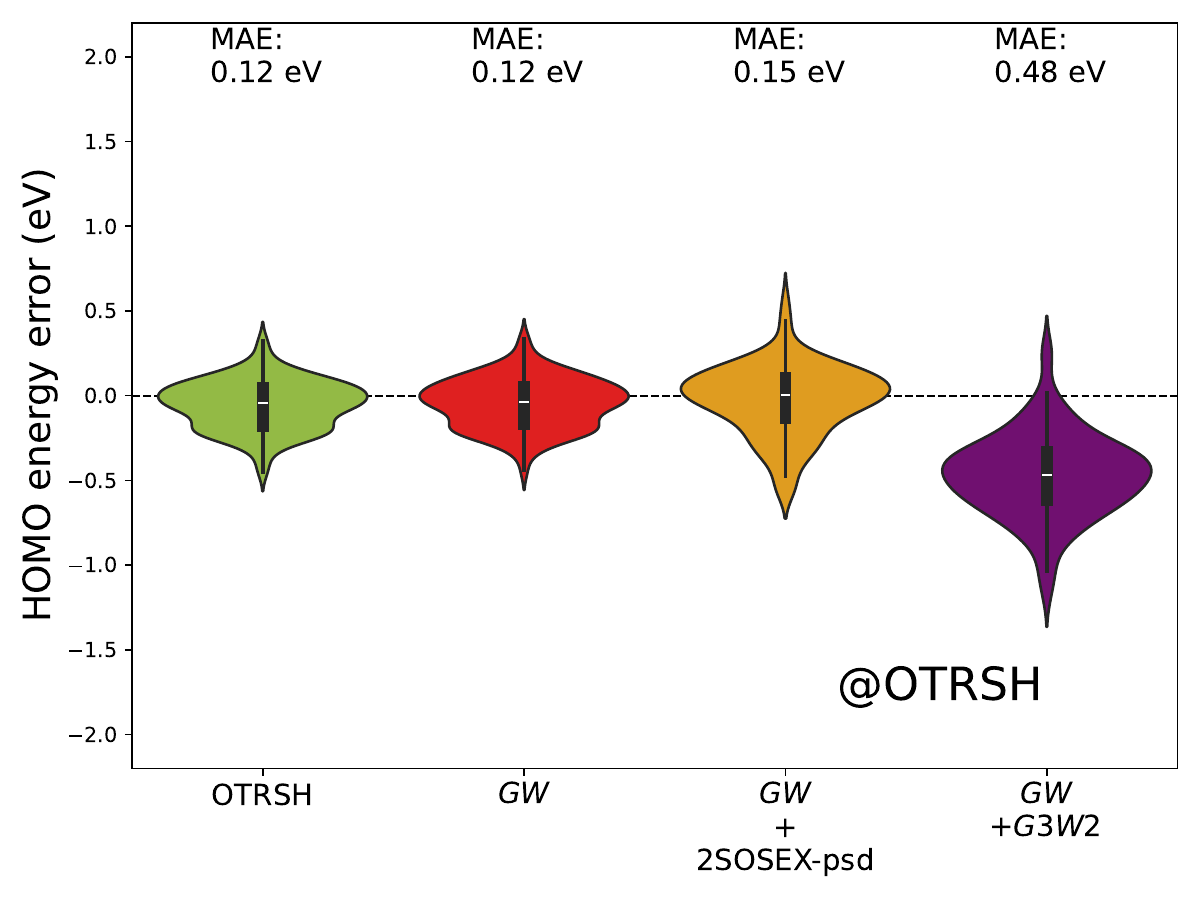}
\caption{
Error distribution of the HOMO energies with respect to a reference CCSD(T) $\Delta$SCF evaluation
for the 100 molecules contained in the GW100 set.
The mean absolute errors (MAE) are given for each self-energy approximation.
Self-energies are evaluated as one-shot perturbations on top of PBEh(0.75) (left-hand panel)
or on top of optimally-tuned range-separated hybrid (OTRSH) (right-hand panel).
}
\label{fig:gw100otrsh}
\end{figure*}

Figure~\ref{fig:gw100otrsh} demonstrates that the above conclusions are preserved when using improved starting points.
Here we propose two mean-field approximations, PBEh(0.75) that was highlighted as a very good $GW$ starting point in Ref.~\citenum{bruneval_fchem2021}
and optimally-tuned range-separated hybrid \cite{refaely_prl2012} (OTRSH) that is also known to be excellent when combined with $GW$ \cite{McKeon2022}.
Both starting points are confirmed to have narrow distributions around zero by Fig.~\ref{fig:gw100otrsh}.
$GW$ is excellent with very low MAE and \psd performs equally well.
Again, the numerically more involved $GW$+$G3W2$ approximation is detrimental to the accuracy.

\subsection{CORE65 benchmark for core electron binding energies}

\begin{figure}[hbt!]
\includegraphics[width=0.45\columnwidth]{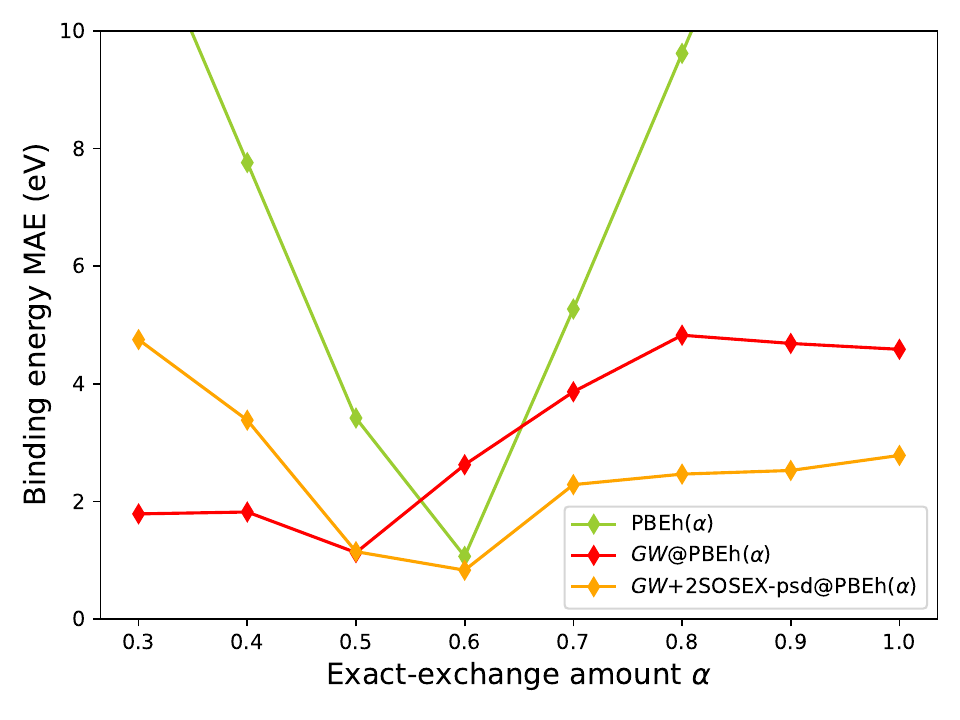}
\caption{
Binding energy mean absolute error for
 a subset of 53 out of 65 core state energies of organic molecules from the CORE65 benchmark
 \cite{golze_jpcl2020}. 
 The MAEs are presented as functions of the amount of exact-exchange in the starting mean-field PBEh($\alpha$).
}
\label{fig:core65}
\end{figure}

The self-energy formulas we derived are analytic and therefore can be evaluated accurately at any frequency.
It is therefore straightforward to consider core ionization energies here, without the need to perform tedious numerical integrations\cite{golze_jctc2018}.

We employ a basis set dedicated to core excitations, pcSseg-3 \cite{jensen_jctc2015}, as prescribed in Ref.~\citenum{mejia_jctc2022}.
To allow for a fair comparison, we used the PBEh($\alpha$) functional with different fractions of exact-exchange as the starting point.
Golze \textit{et al.} \cite{golze_jpcl2020} identified $\alpha=0.45$ as the best starting point for $GW$.
However, this might not be the case for other self-energy approximations.

In Fig.~\ref{fig:core65}, we report the MAE of the binding energies of 53 core states in the CORE65 set \cite{golze_jpcl2020}
as a function of the content of exact-exchange in the starting mean-field.
While we confirm that $GW$ work best with $\sim 0.50$, \psd performs even slightly better for $\alpha=0.6$
which is precisely the PBEh functional that works best.
It is indeed a desirable feature that the accuracy of the final approximation does not rely on a sub-optimal starting point in combination with error compensations.

\section{Conclusion}
\label{sec:conclusion}

Many-body perturbation theory is a standard tool in the arsenal of computational quantum chemistry. Despite years of intensive research, the question of which approximation works best for a given system can only be answered empirically. One approach to make this knowledge more systematic is to restrict to approximations that \emph{a priori} satisfy a number of physical constraints.

This idea has been very fruitful, for example, in the development of exchange-correlation functionals, where incorporating as many known constraints as possible has led to significant accuracy improvements.\cite{Kaplan2023} In many-body perturbation theory, the concept of conserving approximations is another well-known and extensively studied constraint.\cite{almbladh_ijmpb1999} This constraint has frequently been used to support the use of one-shot methods, as partially self-consistent approaches typically do not satisfy conservation laws.\cite{VonBarth1996}

However, there are additional constraints within the framework of many-body perturbation theory. In this work, we focused on the constraint of positive semi-definiteness (PSD) of the electronic spectral function, which corresponds to the same property in the scattering rate function, i.e., the spectral function of the electron self-energy. This condition has a clear physical interpretation: it ensures that the probability of finding an electron at a given energy is always non-negative.

The formalism to construct PSD approximations, i.e., approximations that yield manifestly positive semi-definite spectral functions, has been known for over a decade. It is based on nonequilibrium Green’s function techniques and was initially applied to translationally invariant systems. In this work, we extended the approach to molecular systems. Starting from a subset of second-order diagrams, specifically second-order screened exchange (SOSEX), we observed that this approximation is not PSD. We then constructed a PSD version by including a higher-order diagram. The resulting approximation, which we call $GW$+2SOSEX-psd, not only satisfies the positivity constraint but also leads to a highly efficient numerical scheme with computational cost comparable to that of the analytic approach to the $GW$ approximation. This efficiency is due to our discovery of a cancellation between the static and frequency-dependent components of the screened interaction in certain classes of self-energy diagrams. The effect is pronounced in finite systems and takes place under some conditions at higher orders, as we have demonstrated.

A natural question is how well the new approximation performs relative to existing methods. We conducted extensive numerical tests of ionization potentials and core-level binding energies using the established GW100 and CORE65 benchmarks using the MOLGW code\cite{bruneval_cpc2016}. The main conclusion is that $GW$+2SOSEX-psd consistently outperforms its non-PSD counterparts like $GW$+SOSEX, $GW$+2SOSEX, and $GW$+$G3W2$,  although the improvement over $GW$ itself is modest for valence states and more pronounced for core states.

\begin{acknowledgement}
Part of this work was performed using HPC resources from GENCI–TGCC (Grant 2024-gen6018). 
\end{acknowledgement}

\begin{suppinfo}
Individual results for each molecule in the GW100 and CORE65 benchmarks are provided as Supporting information.
\end{suppinfo}

\appendix

\section{A handy relation}
\label{app:formula}

In this appendix, we demonstrate the relation given in Eq.~(\ref{eq:relation}).
It is not obvious at first sight and was not known to us.
It is valid for RPA energies $\Omega_s$ and RPA eigenvectors $w_s^{mn}$ and $w_s^{ia}$,
where $m, n$ are any MO and where $i, a$ is an occupied-empty pair of MO.

Within RPA, the irreducible polarizability $\chi$ is obtained from the non-interacting polarizability $\chi_0$
with a Dyson-like equation (space indices are omitted):
\begin{equation}
  \chi = \chi_0 + \chi_0 \cdot v \cdot \chi  .
\end{equation}

Let us symmetrize the expression by multiply with $\vsqrt$ on both sides:
\begin{equation}
\label{eq:chisqrt}
  \vsqrt \chi  \vsqrt 
     = \vsqrt \chi_0 \vsqrt
       + ( \vsqrt \chi_0 \vsqrt ) \cdot ( \vsqrt \chi \vsqrt )  .
\end{equation} 

In practice we use an auxiliary basis to represent $\vsqrt \chi_0 \vsqrt$ and  $\vsqrt \chi \vsqrt$.
This is not necessary for the proof but makes the derivation compact and follows closely what we actually have in the code.

Labeling the auxiliary basis with capital letter $P$, $Q$, etc.,  $\vsqrt \chi_0 \vsqrt$ reads
\begin{equation}
  \left( \vsqrt \chi_0 \vsqrt \right)_{PQ}(\omega)
     = \sum_{i a} 2 (P | i a ) \cdot (Q | i a ) 
      \left[
         \frac{1}{\omega  - (\epsilon_a - \epsilon_i) + \icomp \eta}
         -\frac{1}{\omega + (\epsilon_a - \epsilon_i) - \icomp \eta} 
      \right]  ,
\end{equation}
where $(P | i a )$ are 3-center Coulomb integrals.
After solving the Casida equations \cite{casida_book1995, casida_time-dependent_2009,bruneval_cpc2016} in the RPA,
we have an expression for  $\vsqrt \chi \vsqrt$:
\begin{equation}
  \left( \vsqrt \chi \vsqrt \right)_{PQ}(\omega)
     = \sum_{s} w_s^{P} \cdot w_s^{Q} 
      \left[
         \frac{1}{\omega  - \Omega_s + \icomp \eta}
         -\frac{1}{\omega + \Omega_s  - \icomp \eta} 
      \right]  .
\end{equation}
Note that the spin degeneracy factor 2 is included by scaling coefficients $w_s^{P}$ with $\sqrt{2}$.

Introducing the spectral decompositions of   $\vsqrt \chi_0 \vsqrt$
and  $\vsqrt \chi \vsqrt$ in the Dyson-like equation, Eq.~(\ref{eq:chisqrt}),
we obtain
\begin{multline}
\label{eq:dyson_explicit}
 \sum_{s} w_s^{P} \cdot w_s^{Q}
       \left[
         \frac{1}{\omega  - \Omega_s + \icomp \eta}
         -\frac{1}{\omega + \Omega_s  - \icomp \eta}
      \right]
  = \\
   \sum_{ia} 2 (P | i a ) \cdot (Q | i a )
      \left[
         \frac{1}{\omega  - (\epsilon_a - \epsilon_i) + \icomp \eta}
         -\frac{1}{\omega + (\epsilon_a - \epsilon_i) - \icomp \eta}
      \right] \\
  + \sum_R \sum_{ia} 2 (P | i a ) \cdot (R | i a )
  \left[
     \frac{1}{\omega  - (\epsilon_a - \epsilon_i) + \icomp \eta}
     -\frac{1}{\omega + (\epsilon_a - \epsilon_i) - \icomp \eta}
  \right] \\
     \cdot \sum_s
      w_s^{R} \cdot w_s^{Q}
       \left[
         \frac{1}{\omega  - \Omega_s + \icomp \eta}
         -\frac{1}{\omega + \Omega_s  - \icomp \eta}
      \right]  .
\end{multline}

When taking the limit $\omega \to \epsilon_a - \epsilon_i$,
the left-hand side of Eq.~(\ref{eq:dyson_explicit}) smoothly tends to a finite value,
whereas its right-hand side contains diverging terms.

Therefore, the divergent terms in the right-hand side of Eq.~(\ref{eq:dyson_explicit}) should cancel out.
For simplicity, let us assume that $\epsilon_a - \epsilon_i$ energy difference is not degenerate.

Then for each pair $(i, a)$, we have
\begin{multline}
 2 (P | i a ) \cdot (Q | i a )
    + \sum_R  2 (P | i a ) \cdot (R | i a ) \\
        \cdot \sum_s w_s^{R} \cdot w_s^{Q}
        \left[
          \frac{1}{(\epsilon_a - \epsilon_i)  - \Omega_s + \icomp \eta}
          -\frac{1}{(\epsilon_a - \epsilon_i) + \Omega_s  - \icomp \eta}
       \right] = 0 .
\end{multline}
Here we take the limit of vanishing $\eta$, which induces
a real part and an imaginary part by virtue of the Sokhotski–Plemelj formula.
The imaginary part is a sum of delta functions $\delta(\epsilon_a - \epsilon_i \pm  \Omega_s)$.
For finite systems with discrete excitation energies, we can safely assume that the delta function arguments
are always non-zero and then that the imaginary part vanishes.

The remaining real part yields
\begin{equation}
  (Q | i a ) = - \sum_R  (R | i a ) 
        \cdot \sum_s w_s^{R} \cdot w_s^{Q}
          \frac{2 \Omega_s}{(\epsilon_a - \epsilon_i)^2  - \Omega_s^2 }.
\end{equation}
Multiplying by $\vsqrt$ expressed in any pair of MO $(m, n)$, i.e.  $(Q | m n)$,
we finally obtain the desired expression:
\begin{equation}
  (m n | i a ) = \sum_s w_s^{ia} \cdot w_s^{mn}
          \frac{-2 \Omega_s}{(\epsilon_a - \epsilon_i)^2  - \Omega_s^2 },
\end{equation}
which is nothing else but the condition on the time-ordered screened interaction:
\begin{align}
    \left( mn| W^{--}(\epsilon_a - \epsilon_i)|ia\right)=0.\label{eq:W--:cond}
\end{align}
Note that our notation for the matrix elements of the screened interaction is analogous to that of the bare Coulomb interaction; that is, the same index ordering is used. Symmetries $W^{--}(-\w)=W^{--}(\w)$, and $W^{++}(\w)=-\left[W^{--}(\w)\right]^*$ give rise to more conditions.

\section{Another example of a vanishing diagram\label{app:B}}
\begin{figure*}
\includegraphics[scale=1.2]{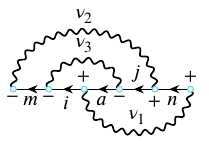} 
\caption{
Example of a vanishing third order partition}
\label{diag:g5w3}
\end{figure*}
Cancellation of diagrams due to the identity~\eqref{eq:relation} also takes places at higher orders. Consider, for instance, one out of six possible third-order SE diagrams, and take a specific ($-$$-$$+$$-$$+$$+$) partition such as illustrated in Fig.~\ref{diag:g5w3} (there are in total $2^4=16$ partitions). Following the diagrammatic rules we obtain
\begin{align*}
 \Sigma^<_{pq}(\w)&=(\icomp)^3\sum_{s,\,miajn}w_s^{pm}w_s^{jn} \int \frac{d\nu_1d\nu_2d\nu_3}{(2\pi)^3}
 \left( ia|W^{++}(\nu_1)|nq\right)  G^{++}_n(\w-\nu_1) G^<_j(\w-\nu_1-\nu_2) W_s^{<}(\nu_2) \\
 &\qquad \times G^>_a(\w-\nu_1-\nu_2-\nu_3) G^<_i(\w-\nu_2-\nu_3)G^{--}_m(\w-\nu_2)
 \left( mi|W^{--}(\nu_3)|aj\right).
\end{align*}
Performing frequency integrals we obtain
\begin{align*}
    \Sigma^<_{pq}(\w)&=-2\pi\icomp\sum_{s,\,miajn}w_s^{pm}w_s^{jn} G^{++}_n(\epsilon_j-\Omega_s)
    G^{--}_m(\epsilon_j-\epsilon_a+\epsilon_i) \\
    &\qquad\left( ia|W^{++}(\epsilon_i-\epsilon_a)|nq\right)
     \left( mi|W^{--}(\epsilon_j-\epsilon_a)|aj\right)\delta(\w-\epsilon_j+\epsilon_a-\epsilon_i+\Omega_s)=0.
\end{align*}
Thus, the partition vanishes in view of the identity~\eqref{eq:W--:cond}. 
\section{Scattering amplitudes involving bare excitations}
After considering the example in Sec.~\ref{app:B}, one may wonder whether diagrammatic expansions in terms of the RPA-screened interaction ever generate terms with bare particle-hole poles. An equivalent statement would be: \emph{A general form of a lesser self-energy expansion contains contributions proportional to 
\begin{align}
    \delta\bigl(\w-\epsilon_j+\sum_{\ell\ge1}\Omega_{\ell}+\sum_{k\in\emptyset}(\epsilon_{a_k}-\epsilon_{i_k})\bigr),
\end{align}
whereas the greater self-energies contain contributions proportional to  
\begin{align}
    \delta\bigl(\w-\epsilon_a-\sum_{\ell\ge1}\Omega_{\ell}-\sum_{k\in\emptyset}(\epsilon_{a_k}-\epsilon_{i_k})\bigr).
\end{align} 
}

\begin{figure*}[t]
\centering
  \begin{tabular}{@{}p{0.3\linewidth}@{\quad}p{0.3\linewidth}@{}}
    \raisebox{1.3cm}{\subfigimg[scale=1.2]{a)}{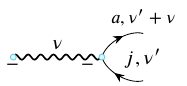}} &
    \subfigimg[scale=1.2]{b)}{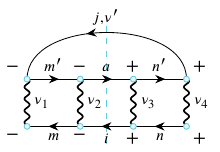} 
  \end{tabular}
\caption{
(a) Time-ordered screened interaction connected to a bare particle–hole excitation. (b) A partition of the fourth-order (in $W$) self-energy diagram with a non-vanishing scattering amplitude that involves the generation of a bare particle–hole excitation.}
\label{diag:vertex}
\end{figure*}

A possible approach for proving such statement would be: Start with a lesser SE term and partition it into two half-diagrams along the $-$/$+$ boundary. For concreteness, consider one of them with vertices on the "$-$"-branch. At the half-diagram boundary, it contains $n+1$ incoming fermionic lines associated with the lesser GFs and $n$ outgoing fermionic lines associated with the greater GFs. On the exit, it contains a single fermionic line. Internally, fermionic lines are connected by interacting lines associated with time-ordered interaction $W^{--}$, so that there is no dangling edges. If there is an interaction line connected to an external particle–hole pair (as shown in Fig.~\ref{diag:vertex}, panel (a)), one can immediately verify that the corresponding contribution vanishes. To see this, let $\nu$ be the frequency associated with $W^{--}$ and $\nu'$ the frequency of the incoming lesser Green’s function, and manipulate the corresponding expression as follows:
\begin{align*}
    \left( mn |W^{--}(\nu)|ai\right) G^<_i(\nu')G^>_a(\nu'+\nu)&=(2\pi)^2W^{--}(\nu)\delta(\nu'-\epsilon_i)\delta(\nu'+\nu-\epsilon_a)\\
    &=(2\pi)^2\left( mn |W^{--}(\nu)|ai\right)\delta(\nu'-\epsilon_i)\delta(\nu+\epsilon_i-\epsilon_a)\\
    &=(2\pi)^2\left( mn |W^{--}(\epsilon_a-\epsilon_i)|ai\right)\delta(\nu'-\epsilon_i)\delta(\nu+\epsilon_i-\epsilon_a)=0.
\end{align*}
In the last step, the identity Eq.~\eqref{eq:W--:cond} is used. Thus, contributions of this type always vanish. One can verify that for diagrams of second and third order such argument always works. This represents a great reduction of the number of partitions. For instance a third-order SE diagram depicted in Fig.~\ref{diag:g5w3} consists in total of 16 partitions, but only 5 of them, namely $-$$-$$-$$-$$-$$+$, $-$$-$$-$$-$$+$$+$, $-$$-$$-$$+$$+$$+$, $-$$-$$+$$+$$+$$+$, $-$$+$$+$$+$$+$$+$, are non-vanishing.

In general, however, the argument does not hold. There are indeed fourth-order examples---such as the one shown in Fig.~\ref{diag:vertex}, panel (b)---that yield a non-vanishing contribution proportional to $\delta(\omega - \epsilon_i + \epsilon_a - \epsilon_j)$, indicating the presence of a bare particle–hole excitation. The diagram was constructed specifically to avoid any $W^{--}$ or $W^{++}$ lines connecting to the particle–hole pairs crossing the border, which makes it impossible to apply the argument discussed above. An explicit expression can be derived using the diagrammatic rules (App.\ref{app:D}):
\begin{align*}
    \Sigma^<_{pq}(\w)&=-(\icomp)^4\sum_{iajmm'nn'} \int{d\nu'}\int \prod_{\alpha=1,4}\frac{d\nu_\alpha}{2\pi}
    G^{++}_{n}(\w-\nu_4) G^{<}_{i}(\w-\nu_4-\nu_3) G^{--}_{m}(\w-\nu_4-\nu_3-\nu_2)\\
     &\quad \times G^{--}_{m'}(\nu'+\nu_1) G^{>}_{a}(\nu'+\nu_1+\nu_2) G^{++}_{n'}(\nu'+\nu_1+\nu_2+\nu_3)G^<_j(\nu')\delta\bigl({\textstyle \sum_{\alpha=1,4}\nu_\alpha}\bigr)\\
     &\quad \times \left( pm|W^{--}(\nu_1)|jm'\right) \left( mi|W^{--}(\nu_2)|am'\right)
     \left( in|W^{++}(\nu_3)|an'\right)\left( nq|W^{++}(\nu_4)jn'\right),
\end{align*}
where the leading sign is due to one fermionic loop. Considering that $G^{\lessgtr}$ are proportional to the $\delta$-functions, and that there is an additional energy conserving condition $\delta\bigl({\textstyle \sum_{\alpha=1,4}\nu_\alpha}\bigr)$, the frequency integrals over $\nu'$, $\nu_1$ and $\nu_3$ immediately yield the overall dependence $2\pi\icomp \sum_{iaj}|X_{iaj}|^2\delta(\w-\e_i+\e_a-\e_j)$. However, the pre-factor $X$ cannot be made zero by virtue of Eq.~\eqref{eq:W--:cond} as it equals to
\begin{align*}
    X_{iaj}&=\sum_{mm'}\int\frac{d\nu}{2\pi}
    G^{--}_{m}(\w-\nu-\e_j+\e_a)G^{--}_{m'}(\e_a-\nu)\\
    &\quad\times \left( pm|W^{--}(\e_a-\e_j-\nu)|jm'\right) \left( mi|W^{--}(\nu)|am'\right).
\end{align*}
Thus, the scattering amplitude associated with the generation of bare particle-hole pairs is, in general, non-vanishing.

\section{Diagrammatic rules\label{app:D}}
In this work we use diagrammatic rules as detailed in Ref.~\citenum{stefanucci_nonequilibrium_2013} chapter 10. For completeness, we include a summary of these rules in this appendix.  Diagrams in bare and screened interactions have the same pre-factors. The pre-factor of a self-energy diagram is $i^n(-1)^l$, where $n$ is the number of interacting lines and $l$ is the number of loops. This yields, for instance, for the 1-ring diagram in Fig.~\ref{diag:pt2} the pre-factor
$i^2(-1)^1=1$, while the SOX diagram has a negative pre-factor due to the absence of loops.

Each vertex in a diagram is associated with a time argument. For physical times and normal time ordering (i.e. on the forward ($-$) branch of the Keldysh contour) we obtain the time-ordered diagrams. 
For general contour times, i.e., on both $-$ and $+$ branches of the Keldysh contour, we obtain diagrams of nonequilibrium Green's function theory.  It is often useful to indicate explicitly the branch  ($\pm$) on which each vertex lies. This labeling helps identify the explicit components of the Green's function $G$ and the screened interaction $W$, such as $G^<\equiv G^{-+}$, or $G^{--}$ representing the time-ordered component, and so on.

For systems at equilibrium, the diagrams are typically evaluated in frequency space, by assigning frequencies to propagators. The sum of frequency arguments of the ingoing lines at each vertex has to be equal to the sum of the frequency arguments of the outgoing lines to satisfy energy conservation. This has been used in Fig.~\ref{diag:pt2}, where some of the frequency arguments have been omitted for brevity. In evaluating each diagram, one has to integrate over all internal vertices, which includes integrations over space, spin, and time/frequency. When integrating over frequencies, each integral contributes a factor $1/(2 \pi)$.

In diagrammatic representations of the lesser self-energy, the outgoing external vertex (typically on the left side of the diagram) carries a minus label, and the incoming vertex carries a plus. Diagrams with assigned labels are referred to as partitions. A single diagram generally corresponds to several partitions; however, not all are physically meaningful. Partitions with isolated islands of pluses or minuses vanish, as demonstrated in Ref.~\citenum{stefanucci_prb2014}.

We do not propose a general scheme for ordering or numbering the partitions for the diagram in Fig.~\ref{diag:vertex}(b). Nevertheless, the other diagrams considered in this work are relatively simple, with vertices arranged along a single line. In such cases, a partition can be indicated by listing the vertex labels from the outgoing ($-$) to the incoming ($+$) vertex. For the SOX diagram, this corresponds to the configuration $-+-+$.

\providecommand{\latin}[1]{#1}
\makeatletter
\providecommand{\doi}
  {\begingroup\let\do\@makeother\dospecials
  \catcode`\{=1 \catcode`\}=2 \doi@aux}
\providecommand{\doi@aux}[1]{\endgroup\texttt{#1}}
\makeatother
\providecommand*\mcitethebibliography{\thebibliography}
\csname @ifundefined\endcsname{endmcitethebibliography}
  {\let\endmcitethebibliography\endthebibliography}{}

\end{document}